\documentclass[aps,prl, twocolumn, superscriptaddress, longbibliography]{revtex4-1}

\usepackage{graphicx}
\usepackage{braket}

\usepackage{hyperref}

\usepackage{float}

\begin{document}

% Use the \preprint command to place your local institutional report
% number in the upper righthand corner of the title page in preprint mode.
% Multiple \preprint commands are allowed.
% Use the 'preprintnumbers' class option to override journal defaults
% to display numbers if necessary
%\preprint{}

%Title of paper
\title{Itinerancy-dependent non-collinear spin textures in SrFeO\textsubscript{3}, CaFeO\textsubscript{3}, and CaFeO\textsubscript{3}/SrFeO\textsubscript{3} heterostructures probed via resonant x-ray scattering}

% repeat the \author .. \affiliation  etc. as needed
% \email, \thanks, \homepage, \altaffiliation all apply to the current
% author. Explanatory text should go in the []'s, actual e-mail
% address or url should go in the {}'s for \email and \homepage.
% Please use the appropriate macro foreach each type of information

% \affiliation command applies to all authors since the last
% \affiliation command. The \affiliation command should follow the
% other information
% \affiliation can be followed by \email, \homepage, \thanks as well.
\author{Paul C. Rogge}
%\homepage[]{Your web page}
%\thanks{}
%\altaffiliation{}
\affiliation{Department of Materials Science and Engineering, Drexel University, Philadelphia, Pennsylvania 19104, USA}

\author{Robert J. Green}
\affiliation{Department of Physics \& Engineering Physics, University of Saskatchewan, Saskatoon, Saskatchewan S7N 5E2, Canada}
\affiliation{Stewart Blusson Quantum Matter Institute, University of British Columbia, Vancouver, British Columbia V6T 1Z4, Canada}
%\email[]{rgreen@physics.ubc.ca}

\author{Ronny Sutarto}
\affiliation{Canadian Light Source, Saskatoon, Saskatchewan S7N 2V3, Canada}

\author{Steven J. May}
%\email[]{smay@coe.drexel.edu}
\affiliation{Department of Materials Science and Engineering, Drexel University, Philadelphia, Pennsylvania 19104, USA}

%Collaboration name if desired (requires use of superscriptaddress
%option in \documentclass). \noaffiliation is required (may also be
%used with the \author command).
%\collaboration can be followed by \email, \homepage, \thanks as well.
%\collaboration{}
%\noaffiliation

\date{\today}

\begin{abstract}
Non-collinear, multi-\textit{q} spin textures can give rise to exotic, topologically protected spin structures such as skyrmions, but the reason for their formation over simple single-\textit{q} structures is not well understood. While lattice frustration and the Dzyaloshinskii-Moriya interaction are known to produce non-collinear spin textures, the role of electron itinerancy in multi-\textit{q} formation is much less studied. Here we investigated the non-collinear, helical spin structures in epitaxial films of the perovskite oxides SrFeO\textsubscript{3} and CaFeO\textsubscript{3} using magnetotransport and resonant soft x-ray magnetic diffraction. Metallic SrFeO\textsubscript{3} exhibits features in its magnetoresistance that are consistent with its recently proposed multi-\textit{q} structure. Additionally, the magnetic Bragg peak of SrFeO\textsubscript{3} measured at the Fe $L$ edge resonance energy asymmetrically broadens with decreasing temperature in its multi-\textit{q} state. In contrast, insulating CaFeO\textsubscript{3} has a symmetric scattering peak with an intensity 10x weaker than SrFeO\textsubscript{3}. Enhanced magnetic scattering at O $K$ edge prepeak energies demonstrates the role of a negative charge transfer energy and the resulting oxygen ligand holes in the magnetic ordering of these ferrates. By measuring magnetic diffraction of CaFeO\textsubscript{3}/SrFeO\textsubscript{3} superlattices with thick CaFeO\textsubscript{3} layers, we find that the CaFeO\textsubscript{3} helical ordering is coherent across 1 unit cell-thick SrFeO\textsubscript{3} layers but not 6 unit cell-thick layers. We conclude that insulating CaFeO\textsubscript{3} supports only a simple single-\textit{q} helical structure in contrast to metallic SrFeO\textsubscript{3} that hosts multi-\textit{q} structures. Our results provide important insight into the role of electron itinerancy in the formation of multi-\textit{q} spin structures.
\end{abstract}

% insert suggested PACS numbers in braces on next line

%\maketitle must follow title, authors, abstract, \pacs, and \keywords
\maketitle

% body of paper here - Use proper section commands
% References should be done using the \cite, \ref, and \label commands
%\section{Introduction}
% Put \label in argument of \section for cross-referencing
%\section{\label{}}
%\subsection{}
%\subsubsection{}

\subsection{Introduction}

Non-collinear spin textures provide a platform for the study of magnetic ordering and exchange interactions beyond conventional ferro and antiferromagnets, as well as have potential application in electronic devices and data storage \cite{Fert_skyrmion_review, Hellman_interface_magnetism_review}. Of particular recent interest are materials that support multi-\textit{q} non-collinear spin structures, where the spin structure is a superposition of multiple non-collinear orderings, denoted by ordering wavevectors, $q$, along different crystallographic directions, which can lead to topologically non-trivial spin textures, such as skyrmions \cite{Rosler_skyrmion_theory, Muhlbauer_skyrmions, Tokura_skyrmion_rev1}. Such multi-\textit{q} states are known to arise from lattice distortions that result in a non-zero Dzyaloshinskii-Moriya (DM) interaction \cite{DZYALOSHINSKY_DM, Moriya_DM, Bak_helical_theory_DM} or from frustration on triangular lattices \cite{multiq_lattice_frustration, Batista_frustration_review}. However, some materials have neither a DM interaction nor a frustrated lattice, yet still exhibit multi-\textit{q} spin textures. These include elemental Nd \cite{Nd_quad_q}, actinide monopnictides such as USb \cite{ROSSAT_Uranium_multiq, USb_multiq}, and, as very recently demonstrated, the cubic perovskite SrFeO\textsubscript{3} \cite{Ishiwata_SFO_multiQ}. The underlying source of the multi-\textit{q} behavior in these materials is unclear but is critical to understand in order to tune the properties of multi-\textit{q} spin textures. Previous theoretical studies, however, point to a third consideration--electron itinerancy--such that coupling between itinerant and localized electrons leads to multi-\textit{q} structures over single-\textit{q} structures \cite{Martin_multiq_itineracy, Ozawa_multiq_itineracy, Hayami_multiq_itineracy_2, Hayami_multiq_itineracy, Hayami_multiq_itineracy_3, Kakehashi_itinerancy_multiq}. 

In order to investigate the role of itinerancy in multi-\textit{q} spin structures, we synthesized epitaxial films of SrFeO\textsubscript{3} and CaFeO\textsubscript{3} and probed their magnetic structure using magnetotransport and resonant soft x-ray magnetic diffraction. While both materials exhibit incommensurate, non-collinear helical spin structures along $\langle111\rangle$ (see Fig. \ref{XRD}(a)) \cite{takeda_SFO_magnetism, Adler_SFO_magnetism, Keimer_SFO_neutron, Mostovoy_SFO_PRL, Ishiwata_SFO_magnetism, Chakraverty_SFO_magnetism, Woodward_CFO, Kawasaki_CFO_first_transport}, SrFeO\textsubscript{3} is metallic whereas CaFeO\textsubscript{3} is insulating (CaFeO\textsubscript{3} exhibits an electronic phase transition ${\sim}$170 K above its N{\'e}el temperature) \cite{MacChesney_SFO, Matsuno_CFO_dispro}. We demonstrate that the SrFeO\textsubscript{3} film exhibits the same complex magnetic phase diagram with distinct helical structures as measured in bulk samples, while further finding that transition temperatures between the magnetic phases more closely resembles Co-doped SrFeO\textsubscript{3}, which we attribute to the moderate tensile strain induced by the substrate. Via resonant soft x-ray magnetic diffraction (RXMD), we show that the magnetic Bragg peaks of CaFeO\textsubscript{3} and SrFeO\textsubscript{3} exhibit significantly different behavior as a function of temperature. We discuss the SrFeO\textsubscript{3} results in context of the recently proposed multi-\textit{q} structures \cite{Ishiwata_SFO_multiQ}, and from this we conclude that our results are consistent with insulating CaFeO\textsubscript{3} supporting a simple multi-domain, single-\textit{q} structure. By synthesizing CaFeO\textsubscript{3}/SrFeO\textsubscript{3} superlattices with different SrFeO\textsubscript{3} layer thicknesses, we find the CaFeO\textsubscript{3} helical structure is coherent through a single unit cell-thick SrFeO\textsubscript{3} layer but is not coherent when the SrFeO\textsubscript{3} layer is increased to 6 unit cells, even though both compounds support helical magnetic structures albeit with slightly different wavevector magnitudes. This result further suggests that the differences in the helical structures of SrFeO\textsubscript{3} and CaFeO\textsubscript{3} are more significant than a simple change to the magnitude of a single \textit{q} wavevector. Our findings point to the importance of carrier itinerancy in multi-\textit{q} spin structures and provide insight into the effect of heterostructuring dissimilar helical structures. 

\subsection{Film Results}

Epitaxial, (001)-oriented SrFeO\textsubscript{3} and CaFeO\textsubscript{3} films were deposited by oxygen-assisted molecular beam epitaxy at ${\sim}$650 $^\circ$C with an oxygen partial pressure of 8x$10^{-6}$ Torr (base pressure 4x$10^{-10}$ Torr). The as-grown films were subsequently annealed in the deposition chamber by heating to ${\sim}$600 $^\circ$C in oxygen plasma (200 Watts, 1x$10^{-5}$ Torr chamber pressure) and then cooled in oxygen plasma by progressively turning down the heater to zero output power over approximately one hour, followed by continued exposure to the plasma for another hour to ensure complete cooling to room temperature \cite{Rogge_PRM, Rogge_CFO_XLD}. Because ferrates are known to lose oxygen over time, prior to all measurements the films were re-annealed in oxygen plasma by the same post-growth process to mitigate oxygen deficiency.

\begin{figure}
\includegraphics{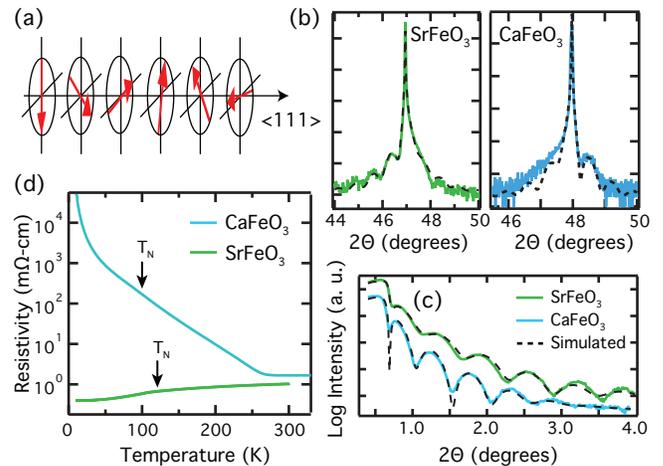}
\caption{(a) Schematic of the incommensurate, single-\textit{q} helical magnetic ordering along one of the four equivalent $\langle111\rangle$ directions. Multi-\textit{q} spin structures arise when the helical ordering exists along different $\langle111\rangle$ directions simultaneously. (b) Hard X-ray diffraction of a SrFeO\textsubscript{3} film deposited on LSAT(001) (0.5\% tensile strain) and a CaFeO\textsubscript{3} film deposited on LaAlO\textsubscript{3}(001) (0.2\% tensile strain) measured at 300 K with $E = 8,047$ eV photons (Cu $K_\alpha$). The simulated diffraction intensity of a perfect epitaxial film is shown by the dashed black line. (c) Hard X-ray reflectivity of the SrFeO\textsubscript{3} and CaFeO\textsubscript{3} films. Simulated intensity (dashed black lines) for a 14.2 nm thick (37 unit cells) SrFeO\textsubscript{3} film and a 15.8 nm thick (42 pseudocubic unit cells) CaFeO\textsubscript{3} film exhibits good agreement with measured. CaFeO\textsubscript{3} data are offset in $y$. (d) Electrical transport of the SrFeO\textsubscript{3} and CaFeO\textsubscript{3} films. Onset of helical ordering is denoted by the arrows (see text).
\label{XRD}}
\end{figure}

SrFeO\textsubscript{3} was deposited on single crystal (La$_{0.18}$Sr$_{0.82}$)(Al$_{0.59}$Ta$_{0.41}$)O$_3$ (LSAT, +0.5\% strain). CaFeO\textsubscript{3} was deposited on LaAlO$_3$ (+0.2\% strain), as were superlattices of SrFeO\textsubscript{3}/CaFeO\textsubscript{3}. X-ray diffraction of the monolithic films is shown in Fig. \ref{XRD}(b). The simulated diffraction intensity of an ideal epitaxial film exhibits good agreement with measured data. Analysis of thickness fringes from x-ray reflectivity measurements shown in Fig. \ref{XRD}(c) indicate that the SrFeO\textsubscript{3} film is 14.2 nm thick (37 unit cells) and the CaFeO\textsubscript{3} film is 15.8 nm thick (42 pseudocubic unit cells), values that approximately match those obtained from the simulated diffraction data. Electrical transport, measured with Ag paint contacts in the van der Pauw geometry, further confirms the high-quality nature of the films. As seen in Fig. \ref{XRD}(d), CaFeO\textsubscript{3} exhibits a metal-insulator transition near 270 K, and both SrFeO\textsubscript{3} and CaFeO\textsubscript{3} have a 300 K resistivity comparable to bulk samples \cite{MacChesney_SFO, Matsuno_CFO_dispro}. The electrical transport confirms that CaFeO\textsubscript{3} is insulating below its N{\'e}el temperature, whereas SrFeO\textsubscript{3} is metallic.

\begin{figure}
 \includegraphics{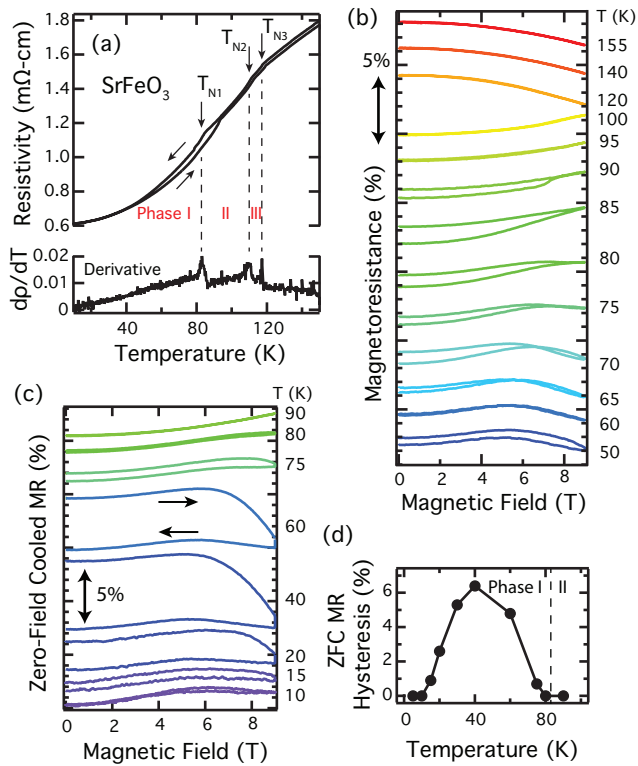}
 \caption{(a) Electrical transport of the SrFeO\textsubscript{3} film exhibits three anomalies as indicated by the spikes in the derivative with temperature (lower panel) and are attributed to helical Phases I, II, and III. (b) Magnetoresistance measurements (H$\parallel$[001], I$\parallel$[100]) of the SrFeO\textsubscript{3} film measured in succession with increasing temperature. The data are shifted in $y$ for clarity. Double arrow scale bar denotes 5\% MR. (c) Zero-field cooled (ZFC) MR measured at various temperatures. The data are shifted in $y$ for clarity. (d) Percent hysteresis in the zero-field cooled MR as a function of temperature. The vertical, dashed line represents the transition between Phase I and II as identified by the electrical resistivity in (a).
 \label{MR}}
 \end{figure}

The electrical transport and magnetoresistance of SrFeO\textsubscript{3} is known to exhibit rich features due to its helical magnetic ordering. Electrical transport, shown in Fig. \ref{MR}(a), exhibits three anomalies as identified by the derivative of the resistivity with temperature (cooling). The first at 117 K is attributed to the onset of magnetic ordering, and a second and third anomaly occur at 110 K and 83 K, respectively. Additionally, the electrical resistivity exhibits hysteresis with temperature. These anomalies and hysteresis are consistent with previous measurements of bulk SrFeO\textsubscript{3}, although a resistivity anomaly at the onset of helical ordering was not observed previously \cite{Lebon_Keimer_SFO, Ishiwata_SFO_magnetism, Long_Co_doped_SFO, Chakraverty_SFO_magnetism}. We adopt the previously used nomenclature and label the regions as Phases I, II, and III \cite{Ishiwata_SFO_magnetism}, as shown in Fig. \ref{MR}(a). No such anomalies are observed in the CaFeO\textsubscript{3} electrical transport \cite{SI_RMD}.

Magnetoresistance [MR $= (\rho(\textrm{H}) - \rho(\textrm{H=0}))/\rho(\textrm{H=0})$] measurements out to 9 T on SrFeO\textsubscript{3} are shown in Fig. \ref{MR}(b) and further confirm the distinct nature of the three identified magnetic phases. The transition from a negative slope (e.g., between 155 K -- 120 K) to a positive slope (e.g., 100 K) is consistent with the onset of helical ordering below 117 K \cite{Ishiwata_SFO_magnetism}. At lower temperatures, the slope of the MR changes sign as the field increases, and a small degree of hysteresis is observed. This change in slope has been attributed to a transition to a field-induced fan- or cone-like helical state (Phases IV and V) \cite{Ishiwata_SFO_magnetism}. Interestingly, this inflection point occurs at lower applied fields compared to previous measurements of both bulk and thin film samples of SrFeO\textsubscript{3}, and instead is more comparable to Co-doped SrFeO\textsubscript{3} \cite{Long_Co_doped_SFO, Chakraverty_SFO_magnetism}. 

In order to further distinguish Phase I and II, zero field cooling (ZFC) MR measurements were performed. After each measurement, the sample was heated to 150 K and then cooled with no applied field. The data are shown in Fig. \ref{MR}(c) and reveal that the ZFC MR hysteresis is a strong function of temperature [hysteresis $=(\rho$\textsubscript{ZFC}(H=0) -- $\rho$\textsubscript{0T,9T,0T}(H=0))/$\rho$\textsubscript{ZFC}(H=0)]. An onset of hysteresis is not observed until ${\sim}75$ K that then increases with decreasing temperature. The ZFC MR hysteresis reaches a maximum near 40 K and vanishes below 20 K (see Fig. \ref{MR}(d)). These results are consistent with previous work that attributed the ZFC MR hysteresis to $H$-induced domain rotation, where Phase I exhibits hysteresis but Phase II, notably, does not \cite{Ishiwata_SFO_magnetism}. The onset of MR hysteresis here at ${\sim}75$ K approximately correlates with the transition from Phase II to Phase I as identified by electrical transport (83 K) in Fig. \ref{MR}(a). Additionally, the reduction in hysteresis at lower temperatures suggests that the critical field for domain rotation increases with decreasing temperature below 40 K and is consistent with previous measurements that showed a critical field of ${\sim}$15 T at 4.2 K \cite{Ishiwata_SFO_magnetism,Ishiwata_SFO_multiQ}. Similar MR measurements for CaFeO\textsubscript{3} were not possible due to its large resistivity below its N{\'e}el temperature (100 K). The main difference between these results and previous work is that the temperature range of Phase II is significantly smaller here (83 - 110 K) compared to bulk SrFeO\textsubscript{3} (56 - 110 K)  \cite{Ishiwata_SFO_magnetism} and previous measurements of a SrFeO\textsubscript{3} film (46 - 104 K) \cite{Chakraverty_SFO_magnetism}. Instead, the Phase II temperature range of our SrFeO\textsubscript{3} film is comparable to Co-doped SrFeO\textsubscript{3}, approximately equivalent to 1\% Co-doping (SrFe\textsubscript{0.99}Co\textsubscript{0.01}O\textsubscript{3}) in both bulk and thin-film samples \cite{Long_Co_doped_SFO, Chakraverty_SFO_magnetism}.

\begin{figure*}
 \includegraphics{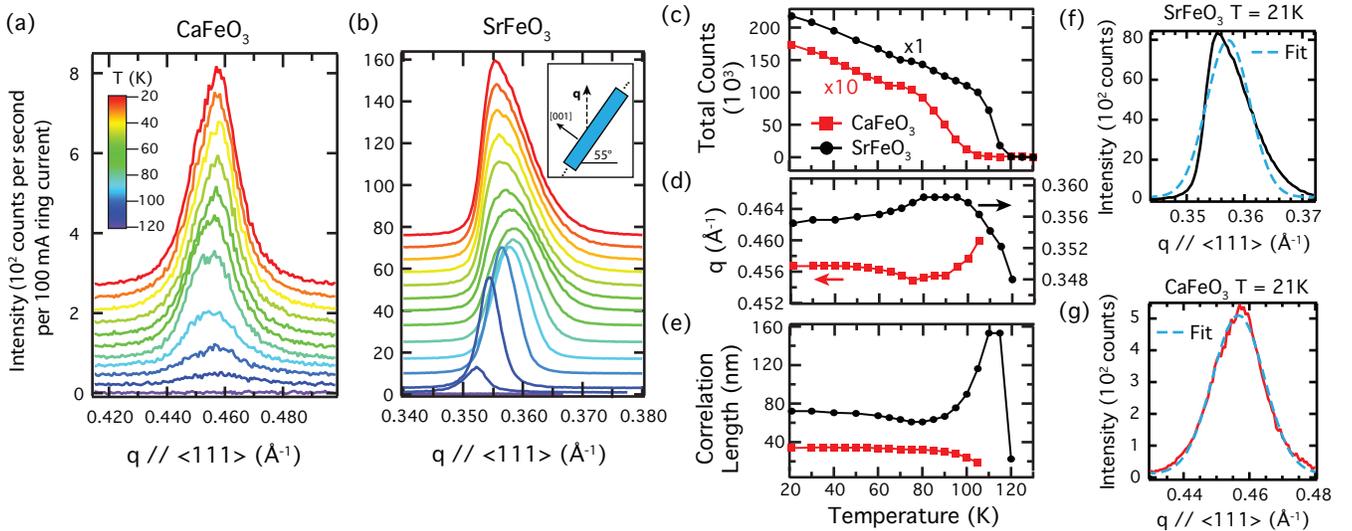}
 \caption{(a) Resonant magnetic scattering along $q_{H=K=L}$ for the CaFeO\textsubscript{3} film (E = 710.8 eV) and (b) the SrFeO\textsubscript{3} film (E = 710.6 eV). Data are offset in $y$. Inset in (b) shows the scattering geometry. (c) Total scattered intensity for the CaFeO\textsubscript{3} and SrFeO\textsubscript{3} films. (d) Magnitude of the scattering vector and (e) correlation length derived from the FWHM of the scattering peak. The FWHM and value of \textit{q} were determined by a Gaussian fit of the CaFeO\textsubscript{3} peaks; for SrFeO\textsubscript{3} the FWHM was manually extracted and \textit{q} was determined by the maximum peak intensity. (f) Gaussian fit (dashed line) of the T = 21 K scattering peak (solid line) for SrFeO\textsubscript{3} and (g) CaFeO\textsubscript{3}.
 \label{scattering}}
 \end{figure*}

To gain further insight, we probed the magnetic ordering by measuring the resonant x-ray scattering at the Fe $L$ edge along $q_{H=K=L}$ for the CaFeO\textsubscript{3} and SrFeO\textsubscript{3} films. Measurements were performed at the REIXS beamline at the Canadian Light Source. The (001)-oriented samples were mounted on a ${\sim}$55 degree wedge in order to place the (111) planes in a symmetric scattering configuration. The scattered intensity measured as a function of temperature for the CaFeO\textsubscript{3} and SrFeO\textsubscript{3} films are shown in Figs. \ref{scattering}(a) and (b), respectively. Both films exhibit scattered intensity at values of $q$ slightly suppressed compared to bulk samples: \textit{q}\textsubscript{CaFeO\textsubscript{3}}$ \sim 0.459$ \AA\textsuperscript{-1} (${\sim}1.37$ nm helical wavelength) (1\% suppressed; bulk: \textit{q}\textsubscript{CaFeO\textsubscript{3}} = 0.465 \AA\textsuperscript{-1} \cite{Woodward_CFO,Kawasaki_CFO_first_transport}); \textit{q}\textsubscript{SrFeO\textsubscript{3}}$ \sim 0.356$ \AA\textsuperscript{-1} (${\sim}1.76$ nm helical wavelength) (3\% suppressed; bulk: \textit{q}\textsubscript{SrFeO\textsubscript{3}} = 0.367 \AA\textsuperscript{-1} \cite{Keimer_SFO_neutron}). The suppressed \textit{q} vector for SrFeO\textsubscript{3} is again equivalent to approximately 1\% Co-doping \cite{Long_Co_doped_SFO}. As seen in Fig. \ref{scattering}(a), the CaFeO\textsubscript{3} peak grows uniformly in intensity with decreasing temperature. Integrating the peak area and plotting as a function of temperature in Fig. \ref{scattering}(c) reveals an onset of magnetic ordering at ${\sim}$100 K, which is slightly suppressed compared to bulk ($T_N = 120$ K \cite{CFO_Neel_temp}). Repeating for the SrFeO\textsubscript{3} film, we find an onset temperature of ${\sim}$115 K, which correlates with $T_{N3} = 117$ K as determined from the electrical transport data, and is similarly suppressed compared to bulk ($T_{N3\textrm{,bulk}} = 133$ K \cite{Keimer_SFO_neutron}). 

From the magnetic scattering data, the \textit{q} vector as a function of temperature was extracted and is plotted in Fig. \ref{scattering}(d). Consistent with previous studies, the SrFeO\textsubscript{3} \textit{q} vector increases in magnitude with decreasing temperature \cite{Keimer_SFO_neutron, Chakraverty_SFO_magnetism}, although here we additionally observe a small decrease below 80 K (Phase II$\rightarrow$I transition). In contrast, CaFeO\textsubscript{3} exhibits the opposite behavior, where the \textit{q} vector decreases with decreasing temperature followed by a small increase at the lowest temperatures. Additionally, the overall change in \textit{q} for CaFeO\textsubscript{3} is approximately 2x smaller than that exhibited by SrFeO\textsubscript{3}.

Comparing the peak shape evolution with temperature between CaFeO\textsubscript{3} and SrFeO\textsubscript{3} in Figs. \ref{scattering}(a) and \ref{scattering}(b), respectively, reveals three striking contrasts. First, SrFeO\textsubscript{3} exhibits a significantly enhanced scattering intensity compared to CaFeO\textsubscript{3}. As seen in Fig. \ref{scattering}(c), the scattered intensity for SrFeO\textsubscript{3} is over 10x greater than CaFeO\textsubscript{3}, which is unexpected given the same nominal film thickness, x-ray footprint, and detector settings (because the \textit{q} vectors are different, the x-ray footprint for CaFeO\textsubscript{3} is reduced by ${\sim}$25\%, which does not sufficiently explain the 10x reduction in intensity). The second observable difference between SrFeO\textsubscript{3} and CaFeO\textsubscript{3} is whereas the CaFeO\textsubscript{3} peak is initially broad and becomes more narrow with decreasing temperature, SrFeO\textsubscript{3} exhibits a narrow peak at the onset of helical ordering that then broadens with decreasing temperature. Converting the FWHM to a correlation length, $\xi = 2\pi/$FWHM, and plotting as a function of temperature in Fig. \ref{scattering}(e) demonstrates that the SrFeO\textsubscript{3} correlation length is initially 4x greater than CaFeO\textsubscript{3} and nearly 2x greater at lower temperatures. We note that the lower bound of the correlation length for an asymmetric Bragg peak of a thin film (e.g., the 111 reflection of a (001)-oriented film) is not limited to the film's thickness.

The third major contrast between SrFeO\textsubscript{3} and CaFeO\textsubscript{3} is the significantly different peak shape evolution exhibited by SrFeO\textsubscript{3}. With decreasing temperature, the peak asymmetrically broadens, where the expansion occurs at higher values of $q$, as highlighted in Fig. \ref{scattering}(f), where a symmetric Gaussian function cannot replicate the peak at 21 K. Such broadening was not observed in previous RXMD measurements of a SrFeO\textsubscript{3} film \cite{Chakraverty_SFO_magnetism}. This broadening ceases below 80 K (see Fig. \ref{scattering}(e)), which correlates with the previously determined Phase II$\rightarrow$I transition in SrFeO\textsubscript{3} ($T_{N1}$ = 83 K) extracted from electrical resistivity and ZFC MR measurements. In contrast, the CaFeO\textsubscript{3} scattering peak at 21 K is replicated by a symmetric Gaussian curve, as shown in Fig. \ref{scattering}(g).

\subsection{Film results discussion}

The helical spin structures of our SrFeO\textsubscript{3} film exhibit features equivalent to approximately 1\% Co-doped SrFeO\textsubscript{3} \cite{Long_Co_doped_SFO, Chakraverty_SFO_magnetism}, as demonstrated by the narrowing of the Phase II temperature window, a decrease in the magnetic field strength at which the MR slope changes sign, a decrease in the magnitude of the \textit{q} vector as probed by RXMD, and a decreased N{\'e}el temperature. A decrease in the \textit{q} vector implies that the real-space length of the helix increases or, analogously, the helical angle between neighboring (111) planes, $\phi$, decreases. A previous theoretical study of the helical state in these ferrates found that the helical ordering arises due to the double exchange effect coupled with a negative charge transfer energy, $\Delta$ \cite{Mostovoy_SFO_PRL, Mostovoy_SFO_2}, where a negative $\Delta$ arises from the high formal oxidation state of Fe\textsuperscript{4+} in these ferrates \cite{Sawatzky_neg_charge_trans_1, Matsuno_CFO_dispro, Robert, Rogge_PRM}. The helical angle tracks a single parameter $\delta = (\epsilon_F - \Delta + t_{pp})/(pd\sigma)$, where $\epsilon_F$ is the Fermi level position, $t_{pp}$ is the oxygen-oxygen hopping amplitude, and $pd\sigma$ is the $\sigma$ hybridization between $p$ and $d$ orbitals. Thus a decrease in $\phi$ corresponds to a decrease in $\delta$, which can occur for an increased $\Delta$, a decreased $t_{pp}$, and/or an increased $pd\sigma$. While tensile strain would be expected to decrease $pd\sigma$, it would also be expected to increase $\Delta$ and decrease $t_{pp}$, which can account for the reduction in $\delta$ and thus a reduction in \textit{q}. In other words, we find that tensile strain acts to increase the ferromagnetic contribution to the SrFeO\textsubscript{3} spin structure, likely due to an increase in $\Delta$ and/or a decrease in $t_{pp}$.

\begin{figure}
 \includegraphics{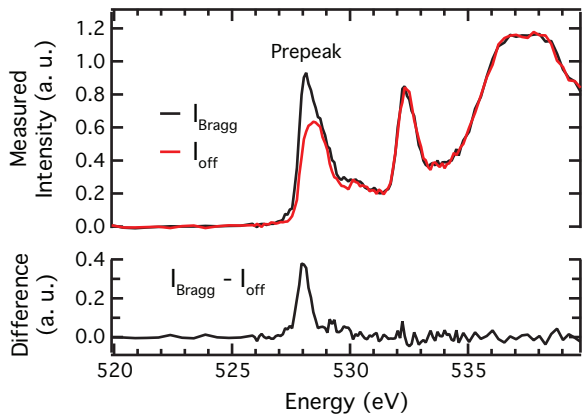}
 \caption{(a) Intensity measured on the magnetic Bragg condition ($I$\textsubscript{Bragg}) and off ($I$\textsubscript{off}) across the O $K$ edge for the SrFeO\textsubscript{3} film on LSAT at 21 K. The O prepeak is at 527.0-529.5 eV, and the peak at 531.5-533.5 eV is from the LSAT substrate. Lower panel shows the magnetic contribution to the scattering as determined by ($I$\textsubscript{Bragg} - $I$\textsubscript{off}).
 \label{Oxy}}
 \end{figure}

To demonstrate that a negative $\Delta$ plays an important role in the magnetic ordering in these ferrates, we measured the resonant magnetic scattering of the SrFeO\textsubscript{3} film as a function of energy across the O $K$ edge on the Bragg condition ($q_z = 0.3548$ \AA\textsuperscript{-1}) and slightly off the Bragg condition (detector offset by 2 degrees). As seen in Fig. \ref{Oxy}, off the Bragg condition the measured intensity has features that resemble an x-ray absorption spectrum of these ferrates, as expected, where the strong prepeak feature between 527.0-529.5 eV arises from the oxygen ligand holes due to the negative charge transfer energy \cite{Abbate_SFO_XAS, Tsuyama_SFO_XPS, Rogge_PRM}. At the Bragg condition, there is a clear enhancement of the intensity within the prepeak region only. Taking the difference between the on and off Bragg conditions isolates the magnetic contribution to the scattered intensity and is shown in the lower panel of Fig. \ref{Oxy}. The observation that the magnetic contribution to the scattered intensity occurs only in the prepeak region supports the conclusion that the charge transfer energy, $\Delta$, is indeed negative and that the interaction between the O $2p$ and Fe $3d$ states is very strong. For completeness, the resonant magnetic scattering across the Fe $L$ edge for the SrFeO\textsubscript{3} and CaFeO\textsubscript{3} films are shown in the Supplemental Material \cite{SI_RMD}.

The resonant magnetic diffraction results for SrFeO\textsubscript{3} correlate with the magnetic phase transitions identified by electrical transport while highlighting distinct differences compared to CaFeO\textsubscript{3}. The evolution of the SrFeO\textsubscript{3} scattering peak is notably unconventional; while the CaFeO\textsubscript{3} correlation length increases slightly upon cooling, as would be expected based on simple thermal considerations, the SrFeO\textsubscript{3} correlation length decreases as a result of asymmetric peak expansion, which we attribute to the transition from Phase II to Phase I upon cooling. In contrast, the CaFeO\textsubscript{3} scattering peak remains symmetric down to at least 21 K and thus suggests that CaFeO\textsubscript{3} does not undergo additional magnetic phase transitions. 

Recently, it has been proposed that SrFeO\textsubscript{3} supports multi-\textit{q} magnetic structures \cite{Ishiwata_SFO_multiQ} such that the spin structure is a superposition of multiple \textit{q} vectors along different crystallographic directions. Specifically, Phase II is proposed to be a single-domain quadruple-\textit{q} structure consisting of proper screw helical ordering with propagation vectors along the four $\langle111\rangle$ vectors of the cubic unit cell \cite{Ishiwata_SFO_multiQ}. The lack of hysteresis in the ZFC MR in Phase II is consistent with a single-domain structure. Additionally, the much larger domain size (correlation length) in Phase II SrFeO\textsubscript{3} compared to CaFeO\textsubscript{3} could arise from a single-domain structure, although we cannot rule out that the CaFeO\textsubscript{3} film may be more defective than the SrFeO\textsubscript{3} film and thus has a smaller domain size.  

The source of the asymmetric change in peak shape and decreased correlation length exhibited by SrFeO\textsubscript{3} at lower temperatures is harder to disentangle with these data alone. We discuss two possible scenarios. In the first scenario, a second, symmetric scattering peak forms at slightly higher \textit{q} values, and grows in intensity with decreasing temperature. Such a scenario would be consistent with Ishiwata \textit{et al.}'s \cite{Ishiwata_SFO_multiQ} proposal of a multi-domain, double-\textit{q} structure in Phase I, where each domain includes both a proper screw ($q_1$) and a vertical cycloid ($q_2$) ordering along one of the four $\langle111\rangle$ directions, where $q_1$ and $q_2$ are along different $\langle111\rangle$ directions, and the magnitude of $q_1$ is approximately equal to the magnitude of $q_2$. In this picture, the two scattering peaks measured along the same [111] direction arise from ordering in different domains: $q_1$ of proper screw along [111] of one domain orientation and $q_2$ of vertical cycloid along [111] of another domain orientation. An additional physical implication for this scenario is that the coherence length is dramatically underestimated because the two distinct peaks were treated as one broad peak. Ishiwata \textit{et al.} also demonstrate that the scattering peak splits into 3 individual peaks upon transition to Phase I from Phase II, where the three peaks are given by ($q$, $q$, $q'$), ($q$, $q'$, $q$), ($q'$, $q$, $q$) and $q'>q$. In a second possible scenario, the asymmetric peak broadening could be due to this splitting, where the precise peak shape depends on the experimental measurement conditions, in particular how \textit{q}-space is scanned and the positions of the three peaks. 

The main takeaway of both scenarios, however, is that the peak shape evolution of SrFeO\textsubscript{3} is not unexpected in context of the recent neutron diffraction measurements and is consistent with a transition from Phase II$\rightarrow$I. Thus our results, particularly the correlation length, highlight how x-ray scattering coupled with thin film effects provides another way to probe multi-\textit{q} magnetic states beyond neutron diffraction measurements. Moreover, these results reveal that the Phase II$\rightarrow$I transition is not as abrupt as indicated in previous phase diagrams, because this transition has been determined based on electrical transport measurements. Here, electrical transport identifies Phase II within 110--83 K, but as seen by the change in the scattering peak shape (as proxied by the correlation length in Fig. \ref{scattering}(e)), the spin structure of SrFeO\textsubscript{3} evolves almost constantly with temperature from 110 to 85 K in Phase II. Thus the single domain Phase II state may  be stable only within a very narrow temperature range just below $T_{N2}$ (110 K). This may account for the lack of an observed decrease in total scattered intensity across the SrFeO\textsubscript{3} Phase II/I transition (single domain to multi-domain structure) because the still rapidly changing total intensity near 110 K could obscure this effect (see Fig. \ref{scattering}(c)).

These results provide important context for analyzing the CaFeO\textsubscript{3} scattering data. While previous studies have identified helical ordering in CaFeO\textsubscript{3} \cite{Woodward_CFO, Kawasaki_CFO_first_transport}, it is unknown if CaFeO\textsubscript{3} supports multi-\textit{q} spin structures, which one may expect given its similarity to SrFeO\textsubscript{3}. However, the data here demonstrate distinct differences between SrFeO\textsubscript{3} and CaFeO\textsubscript{3}. The CaFeO\textsubscript{3} peak grows uniformly with decreasing temperature whereas the SrFeO\textsubscript{3} peak asymmetrically grows and exhibits larger changes in \textit{q}. Additionally, the CaFeO\textsubscript{3} peak at low temperatures is symmetric, as seen in Fig. \ref{scattering}(g), suggesting that it does not replicate the presumed double-\textit{q} ordering seen in SrFeO\textsubscript{3} at low temperatures. Although we cannot definitively determine the precise details of the spin structure within CaFeO\textsubscript{3}, these results are consistent with a multi-domain, single-\textit{q} helical structure, where different domains have helical ordering along one of the four $\langle111\rangle$ directions. A multi-domain helical state in CaFeO\textsubscript{3} is also consistent with the significantly reduced scattering intensity of CaFeO\textsubscript{3} compared to SrFeO\textsubscript{3}--a 4x reduction would be expected given the 4 equivalent $\langle111\rangle$ directions. 

A possible reason for SrFeO\textsubscript{3} hosting multi-\textit{q} spin textures but not CaFeO\textsubscript{3} is their different itinerancies. Previous theoretical studies predict that single-\textit{q} ordering can be destabilized in itinerant systems and instead multi-\textit{q} structures are preferred \cite{Martin_multiq_itineracy, Hayami_multiq_itineracy_2, Hayami_multiq_itineracy, Hayami_multiq_itineracy_3}. While a previous experimental study of the helimagnet Y\textsubscript{3}Co\textsubscript{8}Sn\textsubscript{4} has attributed itinerancy to the source of its multi-\textit{q} state, it also has a DM interaction and lattice frustration that can result in multi-\textit{q} structures \cite{Takagi_multiq_itineracy}. Here, neither lattice frustration nor the DM interaction are present in cubic SrFeO\textsubscript{3}, thus leaving electron itinerancy as the likely source of multi-\textit{q} states in SrFeO\textsubscript{3}. The fact that CaFeO\textsubscript{3} is insulating below its N{\'e}el temperature further supports the conclusion that it has a simple single-\textit{q} spin structure. Future magnetic field-dependent neutron diffraction measurements could confirm a single-\textit{q} structure in CaFeO\textsubscript{3}. Interestingly, tuning the CaFeO\textsubscript{3} metal-insulator transition temperature (e.g., through \textit{A}-site substitution (Ca\textsubscript{1-x}Sr\textsubscript{x}FeO\textsubscript{3} \cite{Takeda_CFO}) or modification of the atomic structure (i.e., octahedral rotations \cite{Antonio_CFO_strain})) below its N{\'e}el temperature could enable further studies of the role of itinerancy in multi-\textit{q} helimagnets. 

\subsection{Superlattice results and discussion}
 
\begin{figure}
 \includegraphics{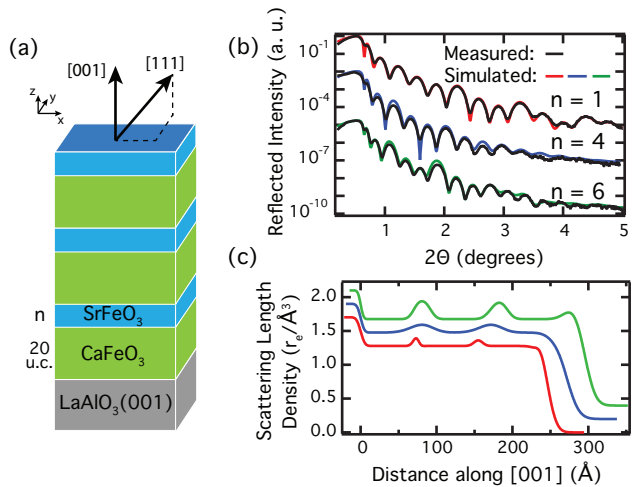}
 \caption{(a) Schematic of the ferrate superlattices studied here. (b) X-ray reflectivity of three ferrate superlattices consisting of [(CaFeO\textsubscript{3})\textsubscript{20}/(SrFeO\textsubscript{3})\textsubscript{n}] x 3 for $n=$1, 4, and 6 unit cells (u. c.) on LaAlO\textsubscript{3}(001) measured at 300 K with $E = 8,047$ eV photons (Cu $K_\alpha$; hard X-rays). Data for $n=4$ and 6 are offset in $y$ for clarity. (c) Real part of the scattering length density corresponding to the simulated reflectivity in (b); data are shifted by 0.2 and 0.4 $r_e$/\AA\textsuperscript{3} for the $n=4$, 6 superlattices, respectively. 
 \label{superlattices_structure}}
 \end{figure}

In order to probe how these different helical structures interact, we synthesized superlattices of CaFeO\textsubscript{3} and SrFeO\textsubscript{3}. Superlattices consisting of [(CaFeO\textsubscript{3})\textsubscript{20}/(SrFeO\textsubscript{3})\textsubscript{n}] x 3 for $n=$ 1, 4, and 6 unit cells were deposited on LaAlO\textsubscript{3}(001), as illustrated in Fig. \ref{superlattices_structure}(a). Non-resonant, hard X-ray reflectivity measurements, shown in Fig. \ref{superlattices_structure}(b), exhibit thickness oscillations and superlattice peaks consistent with the superstructures. The measured and simulated reflectivity for the $n=$ 1, 4, and 6 samples exhibits good agreement, and the corresponding scattering length density for the simulated data, shown in Fig. \ref{superlattices_structure}(c), confirms the superlattice structure. The top SrFeO\textsubscript{3} layer is obscured in the scattering length density for the $n=1$ and 4 superlattices due to the surface roughness (10 \AA{ }and 14 \AA, respectively). All three superlattices exhibit electrical transport similar to the monolithic CaFeO\textsubscript{3} film \cite{SI_RMD}. 

\begin{figure}
 \includegraphics{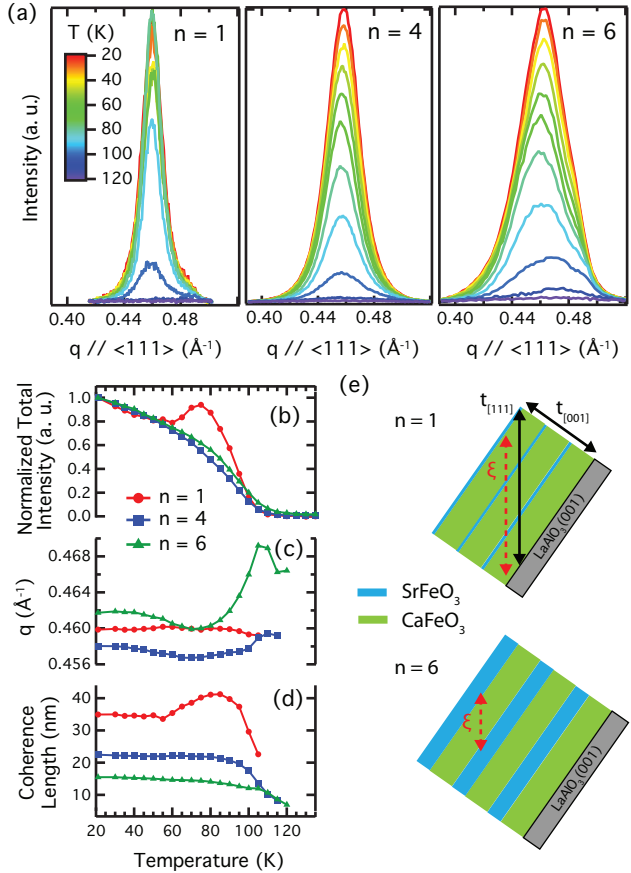}
 \caption{(a) Resonant scattered intensity along $q_{H=K=L}$ for the three superlattices measured at different temperatures with $E = 711.0$ eV. The intensity at 21 K was normalized to unity because different detector settings were used among the superlattices. (b) Total scattered intensity, (c) $q$, and (d) correlation length as a function of temperature. (e) Schematic of the superlattice structure in the angled measurement geometry and depictions of the measured correlation length relative to the relevant film dimensions for the $n=1$ and 6 superlattices.
 \label{superlattices}}
 \end{figure}

The resonant magnetic scattering data obtained from the three superlattices are shown in Fig. \ref{superlattices}(a) and reveal three trends. First, as seen in Fig. \ref{superlattices}(b), the onset temperature is near 115 K for all three superlattices. This is 15 K higher than that measured for the monolithic CaFeO\textsubscript{3} film and is closer to that measured for the SrFeO\textsubscript{3} film ($T_{N3} = 117$ K). The intensity of the $n=1$ superlattice trends differently with temperature compared to the other two superlattices, decreasing below 75 K and increasing again below 60 K. Second, the \textit{q} vector of all three superlattices is approximately equal to that of the monolithic CaFeO\textsubscript{3} film and the trend of the temperature dependent $q$ vector resembles that of the CaFeO\textsubscript{3} film as well.

Third, the correlation length decreases with increasing SrFeO\textsubscript{3} thickness, indicating that the SrFeO\textsubscript{3} layers disrupt the helical ordering. At the lowest temperatures, we find that the coherence length is 37 nm for the $n=1$ superlattice and decreases to 16 nm for $n=6$. For the $n=1$ superlattice, the film thickness along [111] is 41 nm, and in the simple case in which the magnetic domains are isotropic, this would indicate that the helical ordering is coherent through the entire superlattice such that the magnetic moments in the SrFeO\textsubscript{3} layers are ordered coherently with those in the neighboring CaFeO\textsubscript{3} layers. For $n=6$, the 16 nm correlation length is comparable to the thickness of the individual CaFeO\textsubscript{3} layers (${\sim}13$ nm), indicating that the helical ordering within the CaFeO\textsubscript{3} layers does not propagate through the now thicker SrFeO\textsubscript{3} layers. Although measuring down to $q = 0.35$ \AA\textsuperscript{-1}, below the SrFeO\textsubscript{3} wavevector, did not show scattered intensity (not shown), the SrFeO\textsubscript{3} layers may be too thin to detect magnetic ordering, and thus we cannot ascertain definitively if the SrFeO\textsubscript{3} layers of the $n=6$ superlattice are magnetically ordered or not. However, propagation through $n=1$ but not $n=6$ offers further evidence that the helical spin structures of SrFeO\textsubscript{3} and CaFeO\textsubscript{3} are different. Given that both compounds exhibit helical ordering with comparable wavevectors, it is surprising that only 6 SrFeO\textsubscript{3} unit cells disrupts the helical ordering if the spin texture is that of a single \textit{q} helix. The coherence of the spin structure in the $n=1$ superlattice implies that the multi-\textit{q} spin texture of SrFeO\textsubscript{3} has been converted to a single-\textit{q} texture due to proximity to CaFeO\textsubscript{3} and/or confinement effects.

In summary we have explored the role of electron itinerancy in the formation of non-collinear spin textures by studying the magnetic structures of metallic SrFeO\textsubscript{3} and insulating CaFeO\textsubscript{3}. We confirm that our SrFeO\textsubscript{3} film exhibits magnetotransport signatures consistent with its previously determined multi-\textit{q} magnetic structure, and further demonstrate that its resonant soft x-ray magnetic diffraction behavior with temperature is consistent with a multi-\textit{q} spin structure. CaFeO\textsubscript{3}, on the other hand, is found to exhibit significantly different magnetic diffraction characteristics compared to SrFeO\textsubscript{3}, which we attribute to a single-\textit{q} spin helix in CaFeO\textsubscript{3}. Additionally, by synthesizing CaFeO\textsubscript{3}/SrFeO\textsubscript{3} superlattices, we demonstrated that relatively thin layers (6 unit cells) of SrFeO\textsubscript{3} is sufficient to disrupt spin coherency through the superlattice, further supporting the conclusion that SrFeO\textsubscript{3} and CaFeO\textsubscript{3} have different helical magnetic structures. The lack of a Dzyaloshinskii-Moriya interaction and lattice frustration in cubic SrFeO\textsubscript{3}, and the presence of only a single-\textit{q} helical ordering in insulating CaFeO\textsubscript{3}, supports the conclusion that electron itinerancy plays a critical role in the formation of the multi-\textit{q} spin structures in SrFeO\textsubscript{3}. Thus, tuning electron itinerancy in other non-collinear spin structures can potentially be a path towards controlling multi-\textit{q} spin structures and their topologically non-trivial spin structures.

\begin{acknowledgments}
We thank J. A. Borchers and M. R. Fitzsimmons for useful discussions. PCR and SJM were supported by the Army Research Office, grant number W911NF-15-1-0133, and film synthesis at Drexel utilized deposition instrumentation acquired through an Army Research Office DURIP grant (W911NF-14-1-0493). RJG was supported by the Natural Sciences and Engineering Research Council of Canada. Research described in this paper was performed at the Canadian Light Source, which is supported by the Canada Foundation for Innovation, Natural Sciences and Engineering Research Council of Canada, the University of Saskatchewan, the Government of Saskatchewan, Western Economic Diversification Canada, the National Research Council Canada, and the Canadian Institutes of Health Research.
\end{acknowledgments}

% Create the reference section using BibTeX:
%\bibliography{Oxide_bibliography_bibtex}

\begin{thebibliography}{44}%
\makeatletter
\providecommand \@ifxundefined [1]{%
 \@ifx{#1\undefined}
}%
\providecommand \@ifnum [1]{%
 \ifnum #1\expandafter \@firstoftwo
 \else \expandafter \@secondoftwo
 \fi
}%
\providecommand \@ifx [1]{%
 \ifx #1\expandafter \@firstoftwo
 \else \expandafter \@secondoftwo
 \fi
}%
\providecommand \natexlab [1]{#1}%
\providecommand \enquote  [1]{``#1''}%
\providecommand \bibnamefont  [1]{#1}%
\providecommand \bibfnamefont [1]{#1}%
\providecommand \citenamefont [1]{#1}%
\providecommand \href@noop [0]{\@secondoftwo}%
\providecommand \href [0]{\begingroup \@sanitize@url \@href}%
\providecommand \@href[1]{\@@startlink{#1}\@@href}%
\providecommand \@@href[1]{\endgroup#1\@@endlink}%
\providecommand \@sanitize@url [0]{\catcode `\\12\catcode `\$12\catcode
  `\&12\catcode `\#12\catcode `\^12\catcode `\_12\catcode `\%12\relax}%
\providecommand \@@startlink[1]{}%
\providecommand \@@endlink[0]{}%
\providecommand \url  [0]{\begingroup\@sanitize@url \@url }%
\providecommand \@url [1]{\endgroup\@href {#1}{\urlprefix }}%
\providecommand \urlprefix  [0]{URL }%
\providecommand \Eprint [0]{\href }%
\providecommand \doibase [0]{http://dx.doi.org/}%
\providecommand \selectlanguage [0]{\@gobble}%
\providecommand \bibinfo  [0]{\@secondoftwo}%
\providecommand \bibfield  [0]{\@secondoftwo}%
\providecommand \translation [1]{[#1]}%
\providecommand \BibitemOpen [0]{}%
\providecommand \bibitemStop [0]{}%
\providecommand \bibitemNoStop [0]{.\EOS\space}%
\providecommand \EOS [0]{\spacefactor3000\relax}%
\providecommand \BibitemShut  [1]{\csname bibitem#1\endcsname}%
\let\auto@bib@innerbib\@empty
%</preamble>
\bibitem [{\citenamefont {Fert}\ \emph {et~al.}(2017)\citenamefont {Fert},
  \citenamefont {Reyren},\ and\ \citenamefont {Cros}}]{Fert_skyrmion_review}%
  \BibitemOpen
  \bibfield  {author} {\bibinfo {author} {\bibfnamefont {A.}~\bibnamefont
  {Fert}}, \bibinfo {author} {\bibfnamefont {N.}~\bibnamefont {Reyren}}, \ and\
  \bibinfo {author} {\bibfnamefont {V.}~\bibnamefont {Cros}},\ }\bibfield
  {title} {\enquote {\bibinfo {title} {Magnetic skyrmions: {A}dvances in
  physics and potential applications},}\ }\href
  {https://doi.org/10.1038/natrevmats.2017.31} {\bibfield  {journal} {\bibinfo
  {journal} {Nat. Rev. Mater.}\ }\textbf {\bibinfo {volume} {2}},\ \bibinfo
  {pages} {17031} (\bibinfo {year} {2017})}\BibitemShut {NoStop}%
\bibitem [{\citenamefont {Hellman}\ \emph {et~al.}(2017)\citenamefont
  {Hellman}, \citenamefont {Hoffmann}, \citenamefont {Tserkovnyak},
  \citenamefont {Beach}, \citenamefont {Fullerton}, \citenamefont {Leighton},
  \citenamefont {MacDonald}, \citenamefont {Ralph}, \citenamefont {Arena},
  \citenamefont {D{\"u}rr}, \citenamefont {Fischer}, \citenamefont {Grollier},
  \citenamefont {Heremans}, \citenamefont {Jungwirth}, \citenamefont {Kimel},
  \citenamefont {Koopmans}, \citenamefont {Krivorotov}, \citenamefont {May},
  \citenamefont {Petford-Long}, \citenamefont {Rondinelli}, \citenamefont
  {Samarth}, \citenamefont {Schuller}, \citenamefont {Slavin}, \citenamefont
  {Stiles}, \citenamefont {Tchernyshyov}, \citenamefont {Thiaville},\ and\
  \citenamefont {Zink}}]{Hellman_interface_magnetism_review}%
  \BibitemOpen
  \bibfield  {author} {\bibinfo {author} {\bibfnamefont {F.}~\bibnamefont
  {Hellman}}, \bibinfo {author} {\bibfnamefont {A.}~\bibnamefont {Hoffmann}},
  \bibinfo {author} {\bibfnamefont {Y.}~\bibnamefont {Tserkovnyak}}, \bibinfo
  {author} {\bibfnamefont {G.~S.~D.}\ \bibnamefont {Beach}}, \bibinfo {author}
  {\bibfnamefont {E.~E.}\ \bibnamefont {Fullerton}}, \bibinfo {author}
  {\bibfnamefont {C.}~\bibnamefont {Leighton}}, \bibinfo {author}
  {\bibfnamefont {A.~H.}\ \bibnamefont {MacDonald}}, \bibinfo {author}
  {\bibfnamefont {D.~C.}\ \bibnamefont {Ralph}}, \bibinfo {author}
  {\bibfnamefont {D.~A.}\ \bibnamefont {Arena}}, \bibinfo {author}
  {\bibfnamefont {H.~A.}\ \bibnamefont {D{\"u}rr}}, \bibinfo {author}
  {\bibfnamefont {P.}~\bibnamefont {Fischer}}, \bibinfo {author} {\bibfnamefont
  {J.}~\bibnamefont {Grollier}}, \bibinfo {author} {\bibfnamefont {J.~P.}\
  \bibnamefont {Heremans}}, \bibinfo {author} {\bibfnamefont {T.}~\bibnamefont
  {Jungwirth}}, \bibinfo {author} {\bibfnamefont {A.~V.}\ \bibnamefont
  {Kimel}}, \bibinfo {author} {\bibfnamefont {B.}~\bibnamefont {Koopmans}},
  \bibinfo {author} {\bibfnamefont {I.~N.}\ \bibnamefont {Krivorotov}},
  \bibinfo {author} {\bibfnamefont {S.~J.}\ \bibnamefont {May}}, \bibinfo
  {author} {\bibfnamefont {A.~K.}\ \bibnamefont {Petford-Long}}, \bibinfo
  {author} {\bibfnamefont {J.~M.}\ \bibnamefont {Rondinelli}}, \bibinfo
  {author} {\bibfnamefont {N.}~\bibnamefont {Samarth}}, \bibinfo {author}
  {\bibfnamefont {I.~K.}\ \bibnamefont {Schuller}}, \bibinfo {author}
  {\bibfnamefont {A.~N.}\ \bibnamefont {Slavin}}, \bibinfo {author}
  {\bibfnamefont {M.~D.}\ \bibnamefont {Stiles}}, \bibinfo {author}
  {\bibfnamefont {O.}~\bibnamefont {Tchernyshyov}}, \bibinfo {author}
  {\bibfnamefont {A.}~\bibnamefont {Thiaville}}, \ and\ \bibinfo {author}
  {\bibfnamefont {B.~L.}\ \bibnamefont {Zink}},\ }\bibfield  {title} {\enquote
  {\bibinfo {title} {Interface-induced phenomena in magnetism},}\ }\href
  {\doibase 10.1103/RevModPhys.89.025006} {\bibfield  {journal} {\bibinfo
  {journal} {Rev. Mod. Phys.}\ }\textbf {\bibinfo {volume} {89}},\ \bibinfo
  {pages} {025006} (\bibinfo {year} {2017})}\BibitemShut {NoStop}%
\bibitem [{\citenamefont {R{\"o}{\ss}ler}\ \emph {et~al.}(2006)\citenamefont
  {R{\"o}{\ss}ler}, \citenamefont {Bogdanov},\ and\ \citenamefont
  {Pfleiderer}}]{Rosler_skyrmion_theory}%
  \BibitemOpen
  \bibfield  {author} {\bibinfo {author} {\bibfnamefont {U.~K.}\ \bibnamefont
  {R{\"o}{\ss}ler}}, \bibinfo {author} {\bibfnamefont {A.~N.}\ \bibnamefont
  {Bogdanov}}, \ and\ \bibinfo {author} {\bibfnamefont {C.}~\bibnamefont
  {Pfleiderer}},\ }\bibfield  {title} {\enquote {\bibinfo {title} {Spontaneous
  skyrmion ground states in magnetic metals},}\ }\href {\doibase
  10.1038/nature05056} {\bibfield  {journal} {\bibinfo  {journal} {Nature}\
  }\textbf {\bibinfo {volume} {442}},\ \bibinfo {pages} {797--801} (\bibinfo
  {year} {2006})}\BibitemShut {NoStop}%
\bibitem [{\citenamefont {M{\"u}hlbauer}\ \emph {et~al.}(2009)\citenamefont
  {M{\"u}hlbauer}, \citenamefont {Binz}, \citenamefont {Jonietz}, \citenamefont
  {Pfleiderer}, \citenamefont {Rosch}, \citenamefont {Neubauer}, \citenamefont
  {Georgii},\ and\ \citenamefont {B{\"o}ni}}]{Muhlbauer_skyrmions}%
  \BibitemOpen
  \bibfield  {author} {\bibinfo {author} {\bibfnamefont {S.}~\bibnamefont
  {M{\"u}hlbauer}}, \bibinfo {author} {\bibfnamefont {B.}~\bibnamefont {Binz}},
  \bibinfo {author} {\bibfnamefont {F.}~\bibnamefont {Jonietz}}, \bibinfo
  {author} {\bibfnamefont {C.}~\bibnamefont {Pfleiderer}}, \bibinfo {author}
  {\bibfnamefont {A.}~\bibnamefont {Rosch}}, \bibinfo {author} {\bibfnamefont
  {A.}~\bibnamefont {Neubauer}}, \bibinfo {author} {\bibfnamefont
  {R.}~\bibnamefont {Georgii}}, \ and\ \bibinfo {author} {\bibfnamefont
  {P.}~\bibnamefont {B{\"o}ni}},\ }\bibfield  {title} {\enquote {\bibinfo
  {title} {Skyrmion lattice in a chiral magnet},}\ }\href {\doibase
  10.1126/science.1166767} {\bibfield  {journal} {\bibinfo  {journal}
  {Science}\ }\textbf {\bibinfo {volume} {323}},\ \bibinfo {pages} {915--919}
  (\bibinfo {year} {2009})}\BibitemShut {NoStop}%
\bibitem [{\citenamefont {Nagaosa}\ and\ \citenamefont
  {Tokura}(2013)}]{Tokura_skyrmion_rev1}%
  \BibitemOpen
  \bibfield  {author} {\bibinfo {author} {\bibfnamefont {N.}~\bibnamefont
  {Nagaosa}}\ and\ \bibinfo {author} {\bibfnamefont {Y.}~\bibnamefont
  {Tokura}},\ }\bibfield  {title} {\enquote {\bibinfo {title} {Topological
  properties and dynamics of magnetic skyrmions},}\ }\href
  {https://doi.org/10.1038/nnano.2013.243} {\bibfield  {journal} {\bibinfo
  {journal} {Nat. Nano.}\ }\textbf {\bibinfo {volume} {8}},\ \bibinfo {pages}
  {899} (\bibinfo {year} {2013})}\BibitemShut {NoStop}%
\bibitem [{\citenamefont {Dzyaloshinsky}(1958)}]{DZYALOSHINSKY_DM}%
  \BibitemOpen
  \bibfield  {author} {\bibinfo {author} {\bibfnamefont {I.}~\bibnamefont
  {Dzyaloshinsky}},\ }\bibfield  {title} {\enquote {\bibinfo {title} {A
  thermodynamic theory of ``weak'' ferromagnetism of antiferromagnetics},}\
  }\href {\doibase https://doi.org/10.1016/0022-3697(58)90076-3} {\bibfield
  {journal} {\bibinfo  {journal} {J. Phys. Chem. Sol.}\ }\textbf {\bibinfo
  {volume} {4}},\ \bibinfo {pages} {241 -- 255} (\bibinfo {year}
  {1958})}\BibitemShut {NoStop}%
\bibitem [{\citenamefont {Moriya}(1960)}]{Moriya_DM}%
  \BibitemOpen
  \bibfield  {author} {\bibinfo {author} {\bibfnamefont {T.}~\bibnamefont
  {Moriya}},\ }\bibfield  {title} {\enquote {\bibinfo {title} {Anisotropic
  superexchange interaction and weak ferromagnetism},}\ }\href {\doibase
  10.1103/PhysRev.120.91} {\bibfield  {journal} {\bibinfo  {journal} {Phys.
  Rev.}\ }\textbf {\bibinfo {volume} {120}},\ \bibinfo {pages} {91--98}
  (\bibinfo {year} {1960})}\BibitemShut {NoStop}%
\bibitem [{\citenamefont {Bak}\ and\ \citenamefont
  {Jensen}(1980)}]{Bak_helical_theory_DM}%
  \BibitemOpen
  \bibfield  {author} {\bibinfo {author} {\bibfnamefont {P.}~\bibnamefont
  {Bak}}\ and\ \bibinfo {author} {\bibfnamefont {M.~H.}\ \bibnamefont
  {Jensen}},\ }\bibfield  {title} {\enquote {\bibinfo {title} {Theory of
  helical magnetic structures and phase transitions in {MnSi} and {FeGe}},}\
  }\href {\doibase 10.1088/0022-3719/13/31/002} {\bibfield  {journal} {\bibinfo
   {journal} {J. Phys. C Solid State Phys.}\ }\textbf {\bibinfo {volume}
  {13}},\ \bibinfo {pages} {L881--L885} (\bibinfo {year} {1980})}\BibitemShut
  {NoStop}%
\bibitem [{\citenamefont {Okubo}\ \emph {et~al.}(2012)\citenamefont {Okubo},
  \citenamefont {Chung},\ and\ \citenamefont
  {Kawamura}}]{multiq_lattice_frustration}%
  \BibitemOpen
  \bibfield  {author} {\bibinfo {author} {\bibfnamefont {T.}~\bibnamefont
  {Okubo}}, \bibinfo {author} {\bibfnamefont {S.}~\bibnamefont {Chung}}, \ and\
  \bibinfo {author} {\bibfnamefont {H.}~\bibnamefont {Kawamura}},\ }\bibfield
  {title} {\enquote {\bibinfo {title} {Multiple-$q$ states and the skyrmion
  lattice of the triangular-lattice {H}eisenberg antiferromagnet under magnetic
  fields},}\ }\href {\doibase 10.1103/PhysRevLett.108.017206} {\bibfield
  {journal} {\bibinfo  {journal} {Phys. Rev. Lett.}\ }\textbf {\bibinfo
  {volume} {108}},\ \bibinfo {pages} {017206} (\bibinfo {year}
  {2012})}\BibitemShut {NoStop}%
\bibitem [{\citenamefont {Batista}\ \emph {et~al.}(2016)\citenamefont
  {Batista}, \citenamefont {Lin}, \citenamefont {Hayami},\ and\ \citenamefont
  {Kamiya}}]{Batista_frustration_review}%
  \BibitemOpen
  \bibfield  {author} {\bibinfo {author} {\bibfnamefont {C.~D.}\ \bibnamefont
  {Batista}}, \bibinfo {author} {\bibfnamefont {S.~Z.}\ \bibnamefont {Lin}},
  \bibinfo {author} {\bibfnamefont {S.}~\bibnamefont {Hayami}}, \ and\ \bibinfo
  {author} {\bibfnamefont {Y}~\bibnamefont {Kamiya}},\ }\bibfield  {title}
  {\enquote {\bibinfo {title} {Frustration and chiral orderings in correlated
  electron systems},}\ }\href {\doibase 10.1088/0034-4885/79/8/084504}
  {\bibfield  {journal} {\bibinfo  {journal} {Rep. Prog. Phys.}\ }\textbf
  {\bibinfo {volume} {79}},\ \bibinfo {pages} {084504} (\bibinfo {year}
  {2016})}\BibitemShut {NoStop}%
\bibitem [{\citenamefont {Forgan}\ \emph {et~al.}(1989)\citenamefont {Forgan},
  \citenamefont {Gibbons}, \citenamefont {McEwen},\ and\ \citenamefont
  {Fort}}]{Nd_quad_q}%
  \BibitemOpen
  \bibfield  {author} {\bibinfo {author} {\bibfnamefont {E.~M.}\ \bibnamefont
  {Forgan}}, \bibinfo {author} {\bibfnamefont {E.~P.}\ \bibnamefont {Gibbons}},
  \bibinfo {author} {\bibfnamefont {K.~A.}\ \bibnamefont {McEwen}}, \ and\
  \bibinfo {author} {\bibfnamefont {D.}~\bibnamefont {Fort}},\ }\bibfield
  {title} {\enquote {\bibinfo {title} {Observation of a quadruple-$q$ magnetic
  structure in neodymium},}\ }\href {\doibase 10.1103/PhysRevLett.62.470}
  {\bibfield  {journal} {\bibinfo  {journal} {Phys. Rev. Lett.}\ }\textbf
  {\bibinfo {volume} {62}},\ \bibinfo {pages} {470--473} (\bibinfo {year}
  {1989})}\BibitemShut {NoStop}%
\bibitem [{\citenamefont {Rossat-Mignod}\ \emph {et~al.}(1980)\citenamefont
  {Rossat-Mignod}, \citenamefont {Burlet}, \citenamefont {Quezel},\ and\
  \citenamefont {Vogt}}]{ROSSAT_Uranium_multiq}%
  \BibitemOpen
  \bibfield  {author} {\bibinfo {author} {\bibfnamefont {J.}~\bibnamefont
  {Rossat-Mignod}}, \bibinfo {author} {\bibfnamefont {P.}~\bibnamefont
  {Burlet}}, \bibinfo {author} {\bibfnamefont {S.}~\bibnamefont {Quezel}}, \
  and\ \bibinfo {author} {\bibfnamefont {O.}~\bibnamefont {Vogt}},\ }\bibfield
  {title} {\enquote {\bibinfo {title} {Magnetic ordering in cerium and uranium
  monopnictides},}\ }\href {\doibase
  https://doi.org/10.1016/0378-4363(80)90165-5} {\bibfield  {journal} {\bibinfo
   {journal} {Physica B+C}\ }\textbf {\bibinfo {volume} {102}},\ \bibinfo
  {pages} {237 -- 248} (\bibinfo {year} {1980})}\BibitemShut {NoStop}%
\bibitem [{\citenamefont {Normile}\ \emph {et~al.}(2002)\citenamefont
  {Normile}, \citenamefont {Stirling}, \citenamefont {Mannix}, \citenamefont
  {Lander}, \citenamefont {Wastin}, \citenamefont {Rebizant}, \citenamefont
  {Boudarot}, \citenamefont {Burlet}, \citenamefont {Lebech},\ and\
  \citenamefont {Coburn}}]{USb_multiq}%
  \BibitemOpen
  \bibfield  {author} {\bibinfo {author} {\bibfnamefont {P.~S.}\ \bibnamefont
  {Normile}}, \bibinfo {author} {\bibfnamefont {W.~G.}\ \bibnamefont
  {Stirling}}, \bibinfo {author} {\bibfnamefont {D.}~\bibnamefont {Mannix}},
  \bibinfo {author} {\bibfnamefont {G.~H.}\ \bibnamefont {Lander}}, \bibinfo
  {author} {\bibfnamefont {F.}~\bibnamefont {Wastin}}, \bibinfo {author}
  {\bibfnamefont {J.}~\bibnamefont {Rebizant}}, \bibinfo {author}
  {\bibfnamefont {F.}~\bibnamefont {Boudarot}}, \bibinfo {author}
  {\bibfnamefont {P.}~\bibnamefont {Burlet}}, \bibinfo {author} {\bibfnamefont
  {B.}~\bibnamefont {Lebech}}, \ and\ \bibinfo {author} {\bibfnamefont
  {S.}~\bibnamefont {Coburn}},\ }\bibfield  {title} {\enquote {\bibinfo {title}
  {{(U\textsubscript{1-x}Pu\textsubscript{x})Sb} solid solutions. {I.} magnetic
  configurations},}\ }\href {\doibase 10.1103/PhysRevB.66.014405} {\bibfield
  {journal} {\bibinfo  {journal} {Phys. Rev. B}\ }\textbf {\bibinfo {volume}
  {66}},\ \bibinfo {pages} {014405} (\bibinfo {year} {2002})}\BibitemShut
  {NoStop}%
\bibitem [{\citenamefont {Ishiwata}\ \emph {et~al.}(2018)\citenamefont
  {Ishiwata}, \citenamefont {Nakajima}, \citenamefont {Kim}, \citenamefont
  {Inosov}, \citenamefont {Kanazawa}, \citenamefont {White}, \citenamefont
  {Gavilano}, \citenamefont {Georgii}, \citenamefont {Seemann}, \citenamefont
  {Brandl}, \citenamefont {Manuel}, \citenamefont {Khalyavin}, \citenamefont
  {Seki}, \citenamefont {Tokunaga}, \citenamefont {Kinoshita}, \citenamefont
  {Long}, \citenamefont {Kaneko}, \citenamefont {Taguchi}, \citenamefont
  {Arima}, \citenamefont {Keimer},\ and\ \citenamefont
  {Tokura}}]{Ishiwata_SFO_multiQ}%
  \BibitemOpen
  \bibfield  {author} {\bibinfo {author} {\bibfnamefont {S.}~\bibnamefont
  {Ishiwata}}, \bibinfo {author} {\bibfnamefont {T.}~\bibnamefont {Nakajima}},
  \bibinfo {author} {\bibfnamefont {J.~H.}\ \bibnamefont {Kim}}, \bibinfo
  {author} {\bibfnamefont {D.~S.}\ \bibnamefont {Inosov}}, \bibinfo {author}
  {\bibfnamefont {N.}~\bibnamefont {Kanazawa}}, \bibinfo {author}
  {\bibfnamefont {J.~S.}\ \bibnamefont {White}}, \bibinfo {author}
  {\bibfnamefont {J.~L.}\ \bibnamefont {Gavilano}}, \bibinfo {author}
  {\bibfnamefont {R.}~\bibnamefont {Georgii}}, \bibinfo {author} {\bibfnamefont
  {K.}~\bibnamefont {Seemann}}, \bibinfo {author} {\bibfnamefont
  {G.}~\bibnamefont {Brandl}}, \bibinfo {author} {\bibfnamefont
  {P.}~\bibnamefont {Manuel}}, \bibinfo {author} {\bibfnamefont {D.~D.}\
  \bibnamefont {Khalyavin}}, \bibinfo {author} {\bibfnamefont {S.}~\bibnamefont
  {Seki}}, \bibinfo {author} {\bibfnamefont {Y.}~\bibnamefont {Tokunaga}},
  \bibinfo {author} {\bibfnamefont {M.}~\bibnamefont {Kinoshita}}, \bibinfo
  {author} {\bibfnamefont {Y.~W.}\ \bibnamefont {Long}}, \bibinfo {author}
  {\bibfnamefont {Y.}~\bibnamefont {Kaneko}}, \bibinfo {author} {\bibfnamefont
  {Y.}~\bibnamefont {Taguchi}}, \bibinfo {author} {\bibfnamefont
  {T.}~\bibnamefont {Arima}}, \bibinfo {author} {\bibfnamefont
  {B.}~\bibnamefont {Keimer}}, \ and\ \bibinfo {author} {\bibfnamefont
  {Y.}~\bibnamefont {Tokura}},\ }\bibfield  {title} {\enquote {\bibinfo {title}
  {Emergent topological spin structures in a centrosymmetric cubic
  perovskite},}\ }\href {https://arxiv.org/abs/1806.02309} {\bibfield
  {journal} {\bibinfo  {journal} {Preprint at
  https://arxiv.org/abs/1806.02309}\ } (\bibinfo {year} {2018})}\BibitemShut
  {NoStop}%
\bibitem [{\citenamefont {Martin}\ and\ \citenamefont
  {Batista}(2008)}]{Martin_multiq_itineracy}%
  \BibitemOpen
  \bibfield  {author} {\bibinfo {author} {\bibfnamefont {I.}~\bibnamefont
  {Martin}}\ and\ \bibinfo {author} {\bibfnamefont {C.~D.}\ \bibnamefont
  {Batista}},\ }\bibfield  {title} {\enquote {\bibinfo {title} {Itinerant
  electron-driven chiral magnetic ordering and spontaneous quantum hall effect
  in triangular lattice models},}\ }\href {\doibase
  10.1103/PhysRevLett.101.156402} {\bibfield  {journal} {\bibinfo  {journal}
  {Phys. Rev. Lett.}\ }\textbf {\bibinfo {volume} {101}},\ \bibinfo {pages}
  {156402} (\bibinfo {year} {2008})}\BibitemShut {NoStop}%
\bibitem [{\citenamefont {Ozawa}\ \emph {et~al.}(2016)\citenamefont {Ozawa},
  \citenamefont {Hayami}, \citenamefont {Barros}, \citenamefont {Chern},
  \citenamefont {Motome},\ and\ \citenamefont
  {Batista}}]{Ozawa_multiq_itineracy}%
  \BibitemOpen
  \bibfield  {author} {\bibinfo {author} {\bibfnamefont {R.}~\bibnamefont
  {Ozawa}}, \bibinfo {author} {\bibfnamefont {S.}~\bibnamefont {Hayami}},
  \bibinfo {author} {\bibfnamefont {K.}~\bibnamefont {Barros}}, \bibinfo
  {author} {\bibfnamefont {G.~W.}\ \bibnamefont {Chern}}, \bibinfo {author}
  {\bibfnamefont {Y.}~\bibnamefont {Motome}}, \ and\ \bibinfo {author}
  {\bibfnamefont {C.~D.}\ \bibnamefont {Batista}},\ }\bibfield  {title}
  {\enquote {\bibinfo {title} {Vortex crystals with chiral stripes in itinerant
  magnets},}\ }\href {\doibase 10.7566/JPSJ.85.103703} {\bibfield  {journal}
  {\bibinfo  {journal} {J. Phys. Soc. Jpn.}\ }\textbf {\bibinfo {volume}
  {85}},\ \bibinfo {pages} {103703} (\bibinfo {year} {2016})}\BibitemShut
  {NoStop}%
\bibitem [{\citenamefont {Hayami}\ and\ \citenamefont
  {Motome}(2014)}]{Hayami_multiq_itineracy_2}%
  \BibitemOpen
  \bibfield  {author} {\bibinfo {author} {\bibfnamefont {S.}~\bibnamefont
  {Hayami}}\ and\ \bibinfo {author} {\bibfnamefont {Y.}~\bibnamefont
  {Motome}},\ }\bibfield  {title} {\enquote {\bibinfo {title} {Multiple-$q$
  instability by $(d\ensuremath{-}2)$-dimensional connections of {F}ermi
  surfaces},}\ }\href {\doibase 10.1103/PhysRevB.90.060402} {\bibfield
  {journal} {\bibinfo  {journal} {Phys. Rev. B}\ }\textbf {\bibinfo {volume}
  {90}},\ \bibinfo {pages} {060402(R)} (\bibinfo {year} {2014})}\BibitemShut
  {NoStop}%
\bibitem [{\citenamefont {Hayami}\ \emph {et~al.}(2017)\citenamefont {Hayami},
  \citenamefont {Ozawa},\ and\ \citenamefont
  {Motome}}]{Hayami_multiq_itineracy}%
  \BibitemOpen
  \bibfield  {author} {\bibinfo {author} {\bibfnamefont {S.}~\bibnamefont
  {Hayami}}, \bibinfo {author} {\bibfnamefont {R.}~\bibnamefont {Ozawa}}, \
  and\ \bibinfo {author} {\bibfnamefont {Y.}~\bibnamefont {Motome}},\
  }\bibfield  {title} {\enquote {\bibinfo {title} {Effective
  bilinear-biquadratic model for noncoplanar ordering in itinerant magnets},}\
  }\href {\doibase 10.1103/PhysRevB.95.224424} {\bibfield  {journal} {\bibinfo
  {journal} {Phys. Rev. B}\ }\textbf {\bibinfo {volume} {95}},\ \bibinfo
  {pages} {224424} (\bibinfo {year} {2017})}\BibitemShut {NoStop}%
\bibitem [{\citenamefont {Hayami}\ \emph {et~al.}(2014)\citenamefont {Hayami},
  \citenamefont {Misawa}, \citenamefont {Yamaji},\ and\ \citenamefont
  {Motome}}]{Hayami_multiq_itineracy_3}%
  \BibitemOpen
  \bibfield  {author} {\bibinfo {author} {\bibfnamefont {S.}~\bibnamefont
  {Hayami}}, \bibinfo {author} {\bibfnamefont {T.}~\bibnamefont {Misawa}},
  \bibinfo {author} {\bibfnamefont {Y.}~\bibnamefont {Yamaji}}, \ and\ \bibinfo
  {author} {\bibfnamefont {Y.}~\bibnamefont {Motome}},\ }\bibfield  {title}
  {\enquote {\bibinfo {title} {Three-dimensional {D}irac electrons on a cubic
  lattice with noncoplanar multiple-$q$ order},}\ }\href {\doibase
  10.1103/PhysRevB.89.085124} {\bibfield  {journal} {\bibinfo  {journal} {Phys.
  Rev. B}\ }\textbf {\bibinfo {volume} {89}},\ \bibinfo {pages} {085124}
  (\bibinfo {year} {2014})}\BibitemShut {NoStop}%
\bibitem [{\citenamefont {Kakehashi}\ \emph {et~al.}(2018)\citenamefont
  {Kakehashi}, \citenamefont {Koja}, \citenamefont {Olonbayar},\ and\
  \citenamefont {Miyagi}}]{Kakehashi_itinerancy_multiq}%
  \BibitemOpen
  \bibfield  {author} {\bibinfo {author} {\bibfnamefont {Y.}~\bibnamefont
  {Kakehashi}}, \bibinfo {author} {\bibfnamefont {D.}~\bibnamefont {Koja}},
  \bibinfo {author} {\bibfnamefont {T.}~\bibnamefont {Olonbayar}}, \ and\
  \bibinfo {author} {\bibfnamefont {H.}~\bibnamefont {Miyagi}},\ }\bibfield
  {title} {\enquote {\bibinfo {title} {Multiple helical spin density waves and
  magnetic skyrmions in itinerant electron system},}\ }\href {\doibase
  10.7566/JPSJ.87.094712} {\bibfield  {journal} {\bibinfo  {journal} {J. Phys.
  Soc. Jpn.}\ }\textbf {\bibinfo {volume} {87}},\ \bibinfo {pages} {094712}
  (\bibinfo {year} {2018})}\BibitemShut {NoStop}%
\bibitem [{\citenamefont {Takeda}\ \emph {et~al.}(1972)\citenamefont {Takeda},
  \citenamefont {Yamaguchi},\ and\ \citenamefont
  {Watanabe}}]{takeda_SFO_magnetism}%
  \BibitemOpen
  \bibfield  {author} {\bibinfo {author} {\bibfnamefont {T.}~\bibnamefont
  {Takeda}}, \bibinfo {author} {\bibfnamefont {Y.}~\bibnamefont {Yamaguchi}}, \
  and\ \bibinfo {author} {\bibfnamefont {H.}~\bibnamefont {Watanabe}},\
  }\bibfield  {title} {\enquote {\bibinfo {title} {Magnetic structure of
  {SrFeO\textsubscript{3}}},}\ }\href {\doibase 10.1143/JPSJ.33.967} {\bibfield
   {journal} {\bibinfo  {journal} {J. Phys. Soc. Jpn.}\ }\textbf {\bibinfo
  {volume} {33}},\ \bibinfo {pages} {967--969} (\bibinfo {year}
  {1972})}\BibitemShut {NoStop}%
\bibitem [{\citenamefont {Adler}\ \emph {et~al.}(2006)\citenamefont {Adler},
  \citenamefont {Lebon}, \citenamefont {Damljanovi\ifmmode~\acute{c}\else
  \'{c}\fi{}}, \citenamefont {Ulrich}, \citenamefont {Bernhard}, \citenamefont
  {Boris}, \citenamefont {Maljuk}, \citenamefont {Lin},\ and\ \citenamefont
  {Keimer}}]{Adler_SFO_magnetism}%
  \BibitemOpen
  \bibfield  {author} {\bibinfo {author} {\bibfnamefont {P.}~\bibnamefont
  {Adler}}, \bibinfo {author} {\bibfnamefont {A.}~\bibnamefont {Lebon}},
  \bibinfo {author} {\bibfnamefont {V.}~\bibnamefont
  {Damljanovi\ifmmode~\acute{c}\else \'{c}\fi{}}}, \bibinfo {author}
  {\bibfnamefont {C.}~\bibnamefont {Ulrich}}, \bibinfo {author} {\bibfnamefont
  {C.}~\bibnamefont {Bernhard}}, \bibinfo {author} {\bibfnamefont {A.~V.}\
  \bibnamefont {Boris}}, \bibinfo {author} {\bibfnamefont {A.}~\bibnamefont
  {Maljuk}}, \bibinfo {author} {\bibfnamefont {C.~T.}\ \bibnamefont {Lin}}, \
  and\ \bibinfo {author} {\bibfnamefont {B.}~\bibnamefont {Keimer}},\
  }\bibfield  {title} {\enquote {\bibinfo {title} {Magnetoresistance effects in
  {SrFeO}$_{3\ensuremath{-}\ensuremath{\delta}}$: Dependence on phase
  composition and relation to magnetic and charge order},}\ }\href {\doibase
  10.1103/PhysRevB.73.094451} {\bibfield  {journal} {\bibinfo  {journal} {Phys.
  Rev. B}\ }\textbf {\bibinfo {volume} {73}},\ \bibinfo {pages} {094451}
  (\bibinfo {year} {2006})}\BibitemShut {NoStop}%
\bibitem [{\citenamefont {Reehuis}\ \emph {et~al.}(2012)\citenamefont
  {Reehuis}, \citenamefont {Ulrich}, \citenamefont {Maljuk}, \citenamefont
  {Niedermayer}, \citenamefont {Ouladdiaf}, \citenamefont {Hoser},
  \citenamefont {Hofmann},\ and\ \citenamefont {Keimer}}]{Keimer_SFO_neutron}%
  \BibitemOpen
  \bibfield  {author} {\bibinfo {author} {\bibfnamefont {M.}~\bibnamefont
  {Reehuis}}, \bibinfo {author} {\bibfnamefont {C.}~\bibnamefont {Ulrich}},
  \bibinfo {author} {\bibfnamefont {A.}~\bibnamefont {Maljuk}}, \bibinfo
  {author} {\bibfnamefont {Ch.}\ \bibnamefont {Niedermayer}}, \bibinfo {author}
  {\bibfnamefont {B.}~\bibnamefont {Ouladdiaf}}, \bibinfo {author}
  {\bibfnamefont {A.}~\bibnamefont {Hoser}}, \bibinfo {author} {\bibfnamefont
  {T.}~\bibnamefont {Hofmann}}, \ and\ \bibinfo {author} {\bibfnamefont
  {B.}~\bibnamefont {Keimer}},\ }\bibfield  {title} {\enquote {\bibinfo {title}
  {Neutron diffraction study of spin and charge ordering in
  {SrFeO${}_{3\ensuremath{-}\ensuremath{\delta}}$}},}\ }\href {\doibase
  10.1103/PhysRevB.85.184109} {\bibfield  {journal} {\bibinfo  {journal} {Phys.
  Rev. B}\ }\textbf {\bibinfo {volume} {85}},\ \bibinfo {pages} {184109}
  (\bibinfo {year} {2012})}\BibitemShut {NoStop}%
\bibitem [{\citenamefont {Mostovoy}(2005)}]{Mostovoy_SFO_PRL}%
  \BibitemOpen
  \bibfield  {author} {\bibinfo {author} {\bibfnamefont {M.}~\bibnamefont
  {Mostovoy}},\ }\bibfield  {title} {\enquote {\bibinfo {title} {Helicoidal
  ordering in iron perovskites},}\ }\href {\doibase
  10.1103/PhysRevLett.94.137205} {\bibfield  {journal} {\bibinfo  {journal}
  {Phys. Rev. Lett.}\ }\textbf {\bibinfo {volume} {94}},\ \bibinfo {pages}
  {137205} (\bibinfo {year} {2005})}\BibitemShut {NoStop}%
\bibitem [{\citenamefont {Ishiwata}\ \emph {et~al.}(2011)\citenamefont
  {Ishiwata}, \citenamefont {Tokunaga}, \citenamefont {Kaneko}, \citenamefont
  {Okuyama}, \citenamefont {Tokunaga}, \citenamefont {Wakimoto}, \citenamefont
  {Kakurai}, \citenamefont {Arima}, \citenamefont {Taguchi},\ and\
  \citenamefont {Tokura}}]{Ishiwata_SFO_magnetism}%
  \BibitemOpen
  \bibfield  {author} {\bibinfo {author} {\bibfnamefont {S.}~\bibnamefont
  {Ishiwata}}, \bibinfo {author} {\bibfnamefont {M.}~\bibnamefont {Tokunaga}},
  \bibinfo {author} {\bibfnamefont {Y.}~\bibnamefont {Kaneko}}, \bibinfo
  {author} {\bibfnamefont {D.}~\bibnamefont {Okuyama}}, \bibinfo {author}
  {\bibfnamefont {Y.}~\bibnamefont {Tokunaga}}, \bibinfo {author}
  {\bibfnamefont {S.}~\bibnamefont {Wakimoto}}, \bibinfo {author}
  {\bibfnamefont {K.}~\bibnamefont {Kakurai}}, \bibinfo {author} {\bibfnamefont
  {T.}~\bibnamefont {Arima}}, \bibinfo {author} {\bibfnamefont
  {Y.}~\bibnamefont {Taguchi}}, \ and\ \bibinfo {author} {\bibfnamefont
  {Y.}~\bibnamefont {Tokura}},\ }\bibfield  {title} {\enquote {\bibinfo {title}
  {Versatile helimagnetic phases under magnetic fields in cubic perovskite
  {SrFeO$_3$}},}\ }\href {\doibase 10.1103/PhysRevB.84.054427} {\bibfield
  {journal} {\bibinfo  {journal} {Phys. Rev. B}\ }\textbf {\bibinfo {volume}
  {84}},\ \bibinfo {pages} {054427} (\bibinfo {year} {2011})}\BibitemShut
  {NoStop}%
\bibitem [{\citenamefont {Chakraverty}\ \emph {et~al.}(2013)\citenamefont
  {Chakraverty}, \citenamefont {Matsuda}, \citenamefont {Wadati}, \citenamefont
  {Okamoto}, \citenamefont {Yamasaki}, \citenamefont {Nakao}, \citenamefont
  {Murakami}, \citenamefont {Ishiwata}, \citenamefont {Kawasaki}, \citenamefont
  {Taguchi}, \citenamefont {Tokura},\ and\ \citenamefont
  {Hwang}}]{Chakraverty_SFO_magnetism}%
  \BibitemOpen
  \bibfield  {author} {\bibinfo {author} {\bibfnamefont {S.}~\bibnamefont
  {Chakraverty}}, \bibinfo {author} {\bibfnamefont {T.}~\bibnamefont
  {Matsuda}}, \bibinfo {author} {\bibfnamefont {H.}~\bibnamefont {Wadati}},
  \bibinfo {author} {\bibfnamefont {J.}~\bibnamefont {Okamoto}}, \bibinfo
  {author} {\bibfnamefont {Y.}~\bibnamefont {Yamasaki}}, \bibinfo {author}
  {\bibfnamefont {H.}~\bibnamefont {Nakao}}, \bibinfo {author} {\bibfnamefont
  {Y.}~\bibnamefont {Murakami}}, \bibinfo {author} {\bibfnamefont
  {S.}~\bibnamefont {Ishiwata}}, \bibinfo {author} {\bibfnamefont
  {M.}~\bibnamefont {Kawasaki}}, \bibinfo {author} {\bibfnamefont
  {Y.}~\bibnamefont {Taguchi}}, \bibinfo {author} {\bibfnamefont
  {Y.}~\bibnamefont {Tokura}}, \ and\ \bibinfo {author} {\bibfnamefont {H.~Y.}\
  \bibnamefont {Hwang}},\ }\bibfield  {title} {\enquote {\bibinfo {title}
  {Multiple helimagnetic phases and topological hall effect in epitaxial thin
  films of pristine and {Co-doped} {SrFeO$_3$}},}\ }\href {\doibase
  10.1103/PhysRevB.88.220405} {\bibfield  {journal} {\bibinfo  {journal} {Phys.
  Rev. B}\ }\textbf {\bibinfo {volume} {88}},\ \bibinfo {pages} {220405(R)}
  (\bibinfo {year} {2013})}\BibitemShut {NoStop}%
\bibitem [{\citenamefont {Woodward}\ \emph {et~al.}(2000)\citenamefont
  {Woodward}, \citenamefont {Cox}, \citenamefont {Moshopoulou}, \citenamefont
  {Sleight},\ and\ \citenamefont {Morimoto}}]{Woodward_CFO}%
  \BibitemOpen
  \bibfield  {author} {\bibinfo {author} {\bibfnamefont {P.~M.}\ \bibnamefont
  {Woodward}}, \bibinfo {author} {\bibfnamefont {D.~E.}\ \bibnamefont {Cox}},
  \bibinfo {author} {\bibfnamefont {E.}~\bibnamefont {Moshopoulou}}, \bibinfo
  {author} {\bibfnamefont {A.~W.}\ \bibnamefont {Sleight}}, \ and\ \bibinfo
  {author} {\bibfnamefont {S.}~\bibnamefont {Morimoto}},\ }\bibfield  {title}
  {\enquote {\bibinfo {title} {Structural studies of charge disproportionation
  and magnetic order in {CaFeO}$_3$},}\ }\href {\doibase
  10.1103/PhysRevB.62.844} {\bibfield  {journal} {\bibinfo  {journal} {Phys.
  Rev. B}\ }\textbf {\bibinfo {volume} {62}},\ \bibinfo {pages} {844--855}
  (\bibinfo {year} {2000})}\BibitemShut {NoStop}%
\bibitem [{\citenamefont {Kawasaki}\ \emph {et~al.}(1998)\citenamefont
  {Kawasaki}, \citenamefont {Takano}, \citenamefont {Kanno}, \citenamefont
  {Takeda},\ and\ \citenamefont {Fujimori}}]{Kawasaki_CFO_first_transport}%
  \BibitemOpen
  \bibfield  {author} {\bibinfo {author} {\bibfnamefont {S.}~\bibnamefont
  {Kawasaki}}, \bibinfo {author} {\bibfnamefont {M.}~\bibnamefont {Takano}},
  \bibinfo {author} {\bibfnamefont {R.}~\bibnamefont {Kanno}}, \bibinfo
  {author} {\bibfnamefont {T.}~\bibnamefont {Takeda}}, \ and\ \bibinfo {author}
  {\bibfnamefont {A.}~\bibnamefont {Fujimori}},\ }\bibfield  {title} {\enquote
  {\bibinfo {title} {Phase transitions in {Fe}$^{4+}$ (3$d^4$)-perovskite
  oxides dominated by oxygen-hole character},}\ }\href {\doibase
  10.1143/JPSJ.67.1529} {\bibfield  {journal} {\bibinfo  {journal} {J. Phys.
  Soc. Jpn.}\ }\textbf {\bibinfo {volume} {67}},\ \bibinfo {pages} {1529--1532}
  (\bibinfo {year} {1998})}\BibitemShut {NoStop}%
\bibitem [{\citenamefont {MacChesney}\ \emph {et~al.}(1965)\citenamefont
  {MacChesney}, \citenamefont {Sherwood},\ and\ \citenamefont
  {Potter}}]{MacChesney_SFO}%
  \BibitemOpen
  \bibfield  {author} {\bibinfo {author} {\bibfnamefont {J.~B.}\ \bibnamefont
  {MacChesney}}, \bibinfo {author} {\bibfnamefont {R.~C.}\ \bibnamefont
  {Sherwood}}, \ and\ \bibinfo {author} {\bibfnamefont {J.~F.}\ \bibnamefont
  {Potter}},\ }\bibfield  {title} {\enquote {\bibinfo {title} {Electric and
  magnetic properties of the strontium ferrates},}\ }\href {\doibase
  10.1063/1.1697052} {\bibfield  {journal} {\bibinfo  {journal} {J. Chem.
  Phys.}\ }\textbf {\bibinfo {volume} {43}},\ \bibinfo {pages} {1907--1913}
  (\bibinfo {year} {1965})}\BibitemShut {NoStop}%
\bibitem [{\citenamefont {Matsuno}\ \emph {et~al.}(2002)\citenamefont
  {Matsuno}, \citenamefont {Mizokawa}, \citenamefont {Fujimori}, \citenamefont
  {Takeda}, \citenamefont {Kawasaki},\ and\ \citenamefont
  {Takano}}]{Matsuno_CFO_dispro}%
  \BibitemOpen
  \bibfield  {author} {\bibinfo {author} {\bibfnamefont {J.}~\bibnamefont
  {Matsuno}}, \bibinfo {author} {\bibfnamefont {T.}~\bibnamefont {Mizokawa}},
  \bibinfo {author} {\bibfnamefont {A.}~\bibnamefont {Fujimori}}, \bibinfo
  {author} {\bibfnamefont {Y.}~\bibnamefont {Takeda}}, \bibinfo {author}
  {\bibfnamefont {S.}~\bibnamefont {Kawasaki}}, \ and\ \bibinfo {author}
  {\bibfnamefont {M.}~\bibnamefont {Takano}},\ }\bibfield  {title} {\enquote
  {\bibinfo {title} {Different routes to charge disproportionation in
  perovskite-type {Fe} oxides},}\ }\href {\doibase 10.1103/PhysRevB.66.193103}
  {\bibfield  {journal} {\bibinfo  {journal} {Phys. Rev. B}\ }\textbf {\bibinfo
  {volume} {66}},\ \bibinfo {pages} {193103} (\bibinfo {year}
  {2002})}\BibitemShut {NoStop}%
\bibitem [{\citenamefont {Rogge}\ \emph
  {et~al.}(2018{\natexlab{a}})\citenamefont {Rogge}, \citenamefont
  {Chandrasena}, \citenamefont {Cammarata}, \citenamefont {Green},
  \citenamefont {Shafer}, \citenamefont {Lefler}, \citenamefont {Huon},
  \citenamefont {Arab}, \citenamefont {Arenholz}, \citenamefont {Lee},
  \citenamefont {Lee}, \citenamefont {Nem\ifmmode~\check{s}\else
  \v{s}\fi{}\'ak}, \citenamefont {Rondinelli}, \citenamefont {Gray},\ and\
  \citenamefont {May}}]{Rogge_PRM}%
  \BibitemOpen
  \bibfield  {author} {\bibinfo {author} {\bibfnamefont {P.~C.}\ \bibnamefont
  {Rogge}}, \bibinfo {author} {\bibfnamefont {R.~U.}\ \bibnamefont
  {Chandrasena}}, \bibinfo {author} {\bibfnamefont {A.}~\bibnamefont
  {Cammarata}}, \bibinfo {author} {\bibfnamefont {R.~J.}\ \bibnamefont
  {Green}}, \bibinfo {author} {\bibfnamefont {P.}~\bibnamefont {Shafer}},
  \bibinfo {author} {\bibfnamefont {B.~M.}\ \bibnamefont {Lefler}}, \bibinfo
  {author} {\bibfnamefont {A.}~\bibnamefont {Huon}}, \bibinfo {author}
  {\bibfnamefont {A.}~\bibnamefont {Arab}}, \bibinfo {author} {\bibfnamefont
  {E.}~\bibnamefont {Arenholz}}, \bibinfo {author} {\bibfnamefont {H.~N.}\
  \bibnamefont {Lee}}, \bibinfo {author} {\bibfnamefont {T.~L.}\ \bibnamefont
  {Lee}}, \bibinfo {author} {\bibfnamefont {S.}~\bibnamefont
  {Nem\ifmmode~\check{s}\else \v{s}\fi{}\'ak}}, \bibinfo {author}
  {\bibfnamefont {J.~M.}\ \bibnamefont {Rondinelli}}, \bibinfo {author}
  {\bibfnamefont {A.~X.}\ \bibnamefont {Gray}}, \ and\ \bibinfo {author}
  {\bibfnamefont {S.~J.}\ \bibnamefont {May}},\ }\bibfield  {title} {\enquote
  {\bibinfo {title} {Electronic structure of negative charge transfer
  {CaFeO}$_3$ across the metal-insulator transition},}\ }\href {\doibase
  10.1103/PhysRevMaterials.2.015002} {\bibfield  {journal} {\bibinfo  {journal}
  {Phys. Rev. Mater.}\ }\textbf {\bibinfo {volume} {2}},\ \bibinfo {pages}
  {015002} (\bibinfo {year} {2018}{\natexlab{a}})}\BibitemShut {NoStop}%
\bibitem [{\citenamefont {Rogge}\ \emph
  {et~al.}(2018{\natexlab{b}})\citenamefont {Rogge}, \citenamefont {Green},
  \citenamefont {Shafer}, \citenamefont {Fabbris}, \citenamefont {Barbour},
  \citenamefont {Lefler}, \citenamefont {Arenholz}, \citenamefont {Dean},\ and\
  \citenamefont {May}}]{Rogge_CFO_XLD}%
  \BibitemOpen
  \bibfield  {author} {\bibinfo {author} {\bibfnamefont {P.~C.}\ \bibnamefont
  {Rogge}}, \bibinfo {author} {\bibfnamefont {R.~J.}\ \bibnamefont {Green}},
  \bibinfo {author} {\bibfnamefont {P.}~\bibnamefont {Shafer}}, \bibinfo
  {author} {\bibfnamefont {G.}~\bibnamefont {Fabbris}}, \bibinfo {author}
  {\bibfnamefont {A.~M.}\ \bibnamefont {Barbour}}, \bibinfo {author}
  {\bibfnamefont {B.~M.}\ \bibnamefont {Lefler}}, \bibinfo {author}
  {\bibfnamefont {E.}~\bibnamefont {Arenholz}}, \bibinfo {author}
  {\bibfnamefont {M.~P.~M.}\ \bibnamefont {Dean}}, \ and\ \bibinfo {author}
  {\bibfnamefont {S.~J.}\ \bibnamefont {May}},\ }\bibfield  {title} {\enquote
  {\bibinfo {title} {Inverted orbital polarization in strained correlated oxide
  films},}\ }\href {\doibase 10.1103/PhysRevB.98.201115} {\bibfield  {journal}
  {\bibinfo  {journal} {Phys. Rev. B}\ }\textbf {\bibinfo {volume} {98}},\
  \bibinfo {pages} {201115(R)} (\bibinfo {year}
  {2018}{\natexlab{b}})}\BibitemShut {NoStop}%
\bibitem [{\citenamefont {Lebon}\ \emph {et~al.}(2004)\citenamefont {Lebon},
  \citenamefont {Adler}, \citenamefont {Bernhard}, \citenamefont {Boris},
  \citenamefont {Pimenov}, \citenamefont {Maljuk}, \citenamefont {Lin},
  \citenamefont {Ulrich},\ and\ \citenamefont {Keimer}}]{Lebon_Keimer_SFO}%
  \BibitemOpen
  \bibfield  {author} {\bibinfo {author} {\bibfnamefont {A.}~\bibnamefont
  {Lebon}}, \bibinfo {author} {\bibfnamefont {P.}~\bibnamefont {Adler}},
  \bibinfo {author} {\bibfnamefont {C.}~\bibnamefont {Bernhard}}, \bibinfo
  {author} {\bibfnamefont {A.~V.}\ \bibnamefont {Boris}}, \bibinfo {author}
  {\bibfnamefont {A.~V.}\ \bibnamefont {Pimenov}}, \bibinfo {author}
  {\bibfnamefont {A.}~\bibnamefont {Maljuk}}, \bibinfo {author} {\bibfnamefont
  {C.~T.}\ \bibnamefont {Lin}}, \bibinfo {author} {\bibfnamefont
  {C.}~\bibnamefont {Ulrich}}, \ and\ \bibinfo {author} {\bibfnamefont
  {B.}~\bibnamefont {Keimer}},\ }\bibfield  {title} {\enquote {\bibinfo {title}
  {Magnetism, charge order, and giant magnetoresistance in {SrFeO$_{3-\delta}$}
  single crystals},}\ }\href {\doibase 10.1103/PhysRevLett.92.037202}
  {\bibfield  {journal} {\bibinfo  {journal} {Phys. Rev. Lett.}\ }\textbf
  {\bibinfo {volume} {92}},\ \bibinfo {pages} {037202} (\bibinfo {year}
  {2004})}\BibitemShut {NoStop}%
\bibitem [{\citenamefont {Long}\ \emph {et~al.}(2012)\citenamefont {Long},
  \citenamefont {Kaneko}, \citenamefont {Ishiwata}, \citenamefont {Tokunaga},
  \citenamefont {Matsuda}, \citenamefont {Wadati}, \citenamefont {Tanaka},
  \citenamefont {Shin}, \citenamefont {Tokura},\ and\ \citenamefont
  {Taguchi}}]{Long_Co_doped_SFO}%
  \BibitemOpen
  \bibfield  {author} {\bibinfo {author} {\bibfnamefont {Y.~W.}\ \bibnamefont
  {Long}}, \bibinfo {author} {\bibfnamefont {Y.}~\bibnamefont {Kaneko}},
  \bibinfo {author} {\bibfnamefont {S.}~\bibnamefont {Ishiwata}}, \bibinfo
  {author} {\bibfnamefont {Y.}~\bibnamefont {Tokunaga}}, \bibinfo {author}
  {\bibfnamefont {T.}~\bibnamefont {Matsuda}}, \bibinfo {author} {\bibfnamefont
  {H.}~\bibnamefont {Wadati}}, \bibinfo {author} {\bibfnamefont
  {Y.}~\bibnamefont {Tanaka}}, \bibinfo {author} {\bibfnamefont
  {S.}~\bibnamefont {Shin}}, \bibinfo {author} {\bibfnamefont {Y.}~\bibnamefont
  {Tokura}}, \ and\ \bibinfo {author} {\bibfnamefont {Y.}~\bibnamefont
  {Taguchi}},\ }\bibfield  {title} {\enquote {\bibinfo {title} {Evolution of
  magnetic phases in single crystals of
  {SrFe${}_{1\ensuremath{-}x}$Co${}_{x}$O${}_{3}$} solid solution},}\ }\href
  {\doibase 10.1103/PhysRevB.86.064436} {\bibfield  {journal} {\bibinfo
  {journal} {Phys. Rev. B}\ }\textbf {\bibinfo {volume} {86}},\ \bibinfo
  {pages} {064436} (\bibinfo {year} {2012})}\BibitemShut {NoStop}%
\bibitem [{SI_()}]{SI_RMD}%
  \BibitemOpen
  \href@noop {} {\enquote {\bibinfo {title} {See {Supplemental Material} at
  {[URL]} for the {CaFeO}\textsubscript{3} electrical transport and its
  derivative, resonant magnetic scattering of the {SrFeO}\textsubscript{3} and
  {CaFeO}\textsubscript{3} films across the {Fe} ${L}$ edge, and the electrical
  resistance vs temperature of the three superlattice samples.}}\ }\BibitemShut
  {NoStop}%
\bibitem [{\citenamefont {Kanamaru}\ \emph {et~al.}(1970)\citenamefont
  {Kanamaru}, \citenamefont {Miyamoto}, \citenamefont {Mimura}, \citenamefont
  {Koizumi}, \citenamefont {Shimada}, \citenamefont {Kume},\ and\ \citenamefont
  {Shin}}]{CFO_Neel_temp}%
  \BibitemOpen
  \bibfield  {author} {\bibinfo {author} {\bibfnamefont {F.}~\bibnamefont
  {Kanamaru}}, \bibinfo {author} {\bibfnamefont {H.}~\bibnamefont {Miyamoto}},
  \bibinfo {author} {\bibfnamefont {Y.}~\bibnamefont {Mimura}}, \bibinfo
  {author} {\bibfnamefont {M.}~\bibnamefont {Koizumi}}, \bibinfo {author}
  {\bibfnamefont {M.}~\bibnamefont {Shimada}}, \bibinfo {author} {\bibfnamefont
  {S.}~\bibnamefont {Kume}}, \ and\ \bibinfo {author} {\bibfnamefont
  {S.}~\bibnamefont {Shin}},\ }\bibfield  {title} {\enquote {\bibinfo {title}
  {Synthesis of a new perovskite {CaFeO\textsubscript{3}}},}\ }\href {\doibase
  https://doi.org/10.1016/0025-5408(70)90121-2} {\bibfield  {journal} {\bibinfo
   {journal} {Mater. Res. Bull.}\ }\textbf {\bibinfo {volume} {5}},\ \bibinfo
  {pages} {257 -- 261} (\bibinfo {year} {1970})}\BibitemShut {NoStop}%
\bibitem [{\citenamefont {Azhar}\ and\ \citenamefont
  {Mostovoy}(2017)}]{Mostovoy_SFO_2}%
  \BibitemOpen
  \bibfield  {author} {\bibinfo {author} {\bibfnamefont {M.}~\bibnamefont
  {Azhar}}\ and\ \bibinfo {author} {\bibfnamefont {M.}~\bibnamefont
  {Mostovoy}},\ }\bibfield  {title} {\enquote {\bibinfo {title} {Incommensurate
  spiral order from double-exchange interactions},}\ }\href {\doibase
  10.1103/PhysRevLett.118.027203} {\bibfield  {journal} {\bibinfo  {journal}
  {Phys. Rev. Lett.}\ }\textbf {\bibinfo {volume} {118}},\ \bibinfo {pages}
  {027203} (\bibinfo {year} {2017})}\BibitemShut {NoStop}%
\bibitem [{\citenamefont {Mizokawa}\ \emph {et~al.}(2000)\citenamefont
  {Mizokawa}, \citenamefont {Khomskii},\ and\ \citenamefont
  {Sawatzky}}]{Sawatzky_neg_charge_trans_1}%
  \BibitemOpen
  \bibfield  {author} {\bibinfo {author} {\bibfnamefont {T.}~\bibnamefont
  {Mizokawa}}, \bibinfo {author} {\bibfnamefont {D.~I.}\ \bibnamefont
  {Khomskii}}, \ and\ \bibinfo {author} {\bibfnamefont {G.~A.}\ \bibnamefont
  {Sawatzky}},\ }\bibfield  {title} {\enquote {\bibinfo {title} {Spin and
  charge ordering in self-doped {M}ott insulators},}\ }\href {\doibase
  10.1103/PhysRevB.61.11263} {\bibfield  {journal} {\bibinfo  {journal} {Phys.
  Rev. B}\ }\textbf {\bibinfo {volume} {61}},\ \bibinfo {pages} {11263--11266}
  (\bibinfo {year} {2000})}\BibitemShut {NoStop}%
\bibitem [{\citenamefont {Green}\ \emph {et~al.}(2016)\citenamefont {Green},
  \citenamefont {Haverkort},\ and\ \citenamefont {Sawatzky}}]{Robert}%
  \BibitemOpen
  \bibfield  {author} {\bibinfo {author} {\bibfnamefont {R.~J.}\ \bibnamefont
  {Green}}, \bibinfo {author} {\bibfnamefont {M.~W.}\ \bibnamefont
  {Haverkort}}, \ and\ \bibinfo {author} {\bibfnamefont {G.~A.}\ \bibnamefont
  {Sawatzky}},\ }\bibfield  {title} {\enquote {\bibinfo {title} {Bond
  disproportionation and dynamical charge fluctuations in the perovskite
  rare-earth nickelates},}\ }\href {\doibase 10.1103/PhysRevB.94.195127}
  {\bibfield  {journal} {\bibinfo  {journal} {Phys. Rev. B}\ }\textbf {\bibinfo
  {volume} {94}},\ \bibinfo {pages} {195127} (\bibinfo {year}
  {2016})}\BibitemShut {NoStop}%
\bibitem [{\citenamefont {Abbate}\ \emph {et~al.}(1992)\citenamefont {Abbate},
  \citenamefont {de~Groot}, \citenamefont {Fuggle}, \citenamefont {Fujimori},
  \citenamefont {Strebel}, \citenamefont {Lopez}, \citenamefont {Domke},
  \citenamefont {Kaindl}, \citenamefont {Sawatzky}, \citenamefont {Takano},
  \citenamefont {Takeda}, \citenamefont {Eisaki},\ and\ \citenamefont
  {Uchida}}]{Abbate_SFO_XAS}%
  \BibitemOpen
  \bibfield  {author} {\bibinfo {author} {\bibfnamefont {M.}~\bibnamefont
  {Abbate}}, \bibinfo {author} {\bibfnamefont {F.~M.~F.}\ \bibnamefont
  {de~Groot}}, \bibinfo {author} {\bibfnamefont {J.~C.}\ \bibnamefont
  {Fuggle}}, \bibinfo {author} {\bibfnamefont {A.}~\bibnamefont {Fujimori}},
  \bibinfo {author} {\bibfnamefont {O.}~\bibnamefont {Strebel}}, \bibinfo
  {author} {\bibfnamefont {F.}~\bibnamefont {Lopez}}, \bibinfo {author}
  {\bibfnamefont {M.}~\bibnamefont {Domke}}, \bibinfo {author} {\bibfnamefont
  {G.}~\bibnamefont {Kaindl}}, \bibinfo {author} {\bibfnamefont {G.~A.}\
  \bibnamefont {Sawatzky}}, \bibinfo {author} {\bibfnamefont {M.}~\bibnamefont
  {Takano}}, \bibinfo {author} {\bibfnamefont {Y.}~\bibnamefont {Takeda}},
  \bibinfo {author} {\bibfnamefont {H.}~\bibnamefont {Eisaki}}, \ and\ \bibinfo
  {author} {\bibfnamefont {S.}~\bibnamefont {Uchida}},\ }\bibfield  {title}
  {\enquote {\bibinfo {title} {Controlled-valence properties of
  {La}$_{1-x}${Sr}$_x${FeO}$_3$ and {La}$_{1-x}${Sr}$_x${MnO}$_3$ studied by
  soft-x-ray absorption spectroscopy},}\ }\href {\doibase
  10.1103/PhysRevB.46.4511} {\bibfield  {journal} {\bibinfo  {journal} {Phys.
  Rev. B}\ }\textbf {\bibinfo {volume} {46}},\ \bibinfo {pages} {4511--4519}
  (\bibinfo {year} {1992})}\BibitemShut {NoStop}%
\bibitem [{\citenamefont {Tsuyama}\ \emph {et~al.}(2015)\citenamefont
  {Tsuyama}, \citenamefont {Matsuda}, \citenamefont {Chakraverty},
  \citenamefont {Okamoto}, \citenamefont {Ikenaga}, \citenamefont {Tanaka},
  \citenamefont {Mizokawa}, \citenamefont {Hwang}, \citenamefont {Tokura},\
  and\ \citenamefont {Wadati}}]{Tsuyama_SFO_XPS}%
  \BibitemOpen
  \bibfield  {author} {\bibinfo {author} {\bibfnamefont {T.}~\bibnamefont
  {Tsuyama}}, \bibinfo {author} {\bibfnamefont {T.}~\bibnamefont {Matsuda}},
  \bibinfo {author} {\bibfnamefont {S.}~\bibnamefont {Chakraverty}}, \bibinfo
  {author} {\bibfnamefont {J.}~\bibnamefont {Okamoto}}, \bibinfo {author}
  {\bibfnamefont {E.}~\bibnamefont {Ikenaga}}, \bibinfo {author} {\bibfnamefont
  {A.}~\bibnamefont {Tanaka}}, \bibinfo {author} {\bibfnamefont
  {T.}~\bibnamefont {Mizokawa}}, \bibinfo {author} {\bibfnamefont {H.~Y.}\
  \bibnamefont {Hwang}}, \bibinfo {author} {\bibfnamefont {Y.}~\bibnamefont
  {Tokura}}, \ and\ \bibinfo {author} {\bibfnamefont {H.}~\bibnamefont
  {Wadati}},\ }\bibfield  {title} {\enquote {\bibinfo {title} {X-ray
  spectroscopic study of {BaFeO\textsubscript{3}} thin films: {An}
  {Fe}\textsuperscript{4+} ferromagnetic insulator},}\ }\href {\doibase
  10.1103/PhysRevB.91.115101} {\bibfield  {journal} {\bibinfo  {journal} {Phys.
  Rev. B}\ }\textbf {\bibinfo {volume} {91}},\ \bibinfo {pages} {115101}
  (\bibinfo {year} {2015})}\BibitemShut {NoStop}%
\bibitem [{\citenamefont {Takagi}\ \emph {et~al.}(2018)\citenamefont {Takagi},
  \citenamefont {White}, \citenamefont {Hayami}, \citenamefont {Arita},
  \citenamefont {Honecker}, \citenamefont {R{\o}nnow}, \citenamefont {Tokura},\
  and\ \citenamefont {Seki}}]{Takagi_multiq_itineracy}%
  \BibitemOpen
  \bibfield  {author} {\bibinfo {author} {\bibfnamefont {R.}~\bibnamefont
  {Takagi}}, \bibinfo {author} {\bibfnamefont {J.~S.}\ \bibnamefont {White}},
  \bibinfo {author} {\bibfnamefont {S.}~\bibnamefont {Hayami}}, \bibinfo
  {author} {\bibfnamefont {R.}~\bibnamefont {Arita}}, \bibinfo {author}
  {\bibfnamefont {D.}~\bibnamefont {Honecker}}, \bibinfo {author}
  {\bibfnamefont {H.~M.}\ \bibnamefont {R{\o}nnow}}, \bibinfo {author}
  {\bibfnamefont {Y.}~\bibnamefont {Tokura}}, \ and\ \bibinfo {author}
  {\bibfnamefont {S.}~\bibnamefont {Seki}},\ }\bibfield  {title} {\enquote
  {\bibinfo {title} {Multiple-q noncollinear magnetism in an itinerant
  hexagonal magnet},}\ }\href {\doibase 10.1126/sciadv.aau3402} {\bibfield
  {journal} {\bibinfo  {journal} {Sci. Adv.}\ }\textbf {\bibinfo {volume}
  {4}},\ \bibinfo {pages} {eaau3402} (\bibinfo {year} {2018})}\BibitemShut
  {NoStop}%
\bibitem [{\citenamefont {Takeda}\ \emph {et~al.}(2000)\citenamefont {Takeda},
  \citenamefont {Kanno}, \citenamefont {Kawamoto}, \citenamefont {Takano},
  \citenamefont {Kawasaki}, \citenamefont {Kamiyama},\ and\ \citenamefont
  {Izumi}}]{Takeda_CFO}%
  \BibitemOpen
  \bibfield  {author} {\bibinfo {author} {\bibfnamefont {T.}~\bibnamefont
  {Takeda}}, \bibinfo {author} {\bibfnamefont {R.}~\bibnamefont {Kanno}},
  \bibinfo {author} {\bibfnamefont {Y.}~\bibnamefont {Kawamoto}}, \bibinfo
  {author} {\bibfnamefont {M.}~\bibnamefont {Takano}}, \bibinfo {author}
  {\bibfnamefont {S.}~\bibnamefont {Kawasaki}}, \bibinfo {author}
  {\bibfnamefont {T.}~\bibnamefont {Kamiyama}}, \ and\ \bibinfo {author}
  {\bibfnamefont {F.}~\bibnamefont {Izumi}},\ }\bibfield  {title} {\enquote
  {\bibinfo {title} {Metal-semiconductor transition, charge disproportionation,
  and low-temperature structure of
  {Ca\textsubscript{1-x}Sr\textsubscript{x}FeO\textsubscript{3}} synthesized
  under high-oxygen pressure},}\ }\href {\doibase
  10.1016/S1293-2558(00)01088-8} {\bibfield  {journal} {\bibinfo  {journal}
  {Solid State Sci.}\ }\textbf {\bibinfo {volume} {2}},\ \bibinfo {pages} {673
  -- 687} (\bibinfo {year} {2000})}\BibitemShut {NoStop}%
\bibitem [{\citenamefont {Cammarata}\ and\ \citenamefont
  {Rondinelli}(2013)}]{Antonio_CFO_strain}%
  \BibitemOpen
  \bibfield  {author} {\bibinfo {author} {\bibfnamefont {A.}~\bibnamefont
  {Cammarata}}\ and\ \bibinfo {author} {\bibfnamefont {J.~M.}\ \bibnamefont
  {Rondinelli}},\ }\bibfield  {title} {\enquote {\bibinfo {title} {Octahedral
  engineering of orbital polarizations in charge transfer oxides},}\ }\href
  {\doibase 10.1103/PhysRevB.87.155135} {\bibfield  {journal} {\bibinfo
  {journal} {Phys. Rev. B}\ }\textbf {\bibinfo {volume} {87}},\ \bibinfo
  {pages} {155135} (\bibinfo {year} {2013})}\BibitemShut {NoStop}%
\end{thebibliography}

%merlin.mbs apsrev4-1.bst 2010-07-25 4.21a (PWD, AO, DPC) hacked
%Control: key (0)
%Control: author (0) dotless jnrlst
%Control: editor formatted (1) identically to author
%Control: production of article title (0) allowed
%Control: page (1) range
%Control: year (0) verbatim
%Control: production of eprint (0) enabled
%

\end{document}